\documentclass[a4paper,11pt]{article}
\pdfoutput=1 

\usepackage{jcappub} 
\usepackage{ulem}
\usepackage{subfigure}
\usepackage{amssymb}
\usepackage[T1]{fontenc} 
\DeclareMathOperator{\sech}{sech}

\title{\boldmath Resonant instability of axionic dark matter clumps}

\author[a,b]{Zihang Wang,}
\author[b]{Lijing Shao}
\author[b]{and Li-Xin Li}

\affiliation[a]{Department of Astronomy, School of Physics, Peking
University, \\ Beijing 100871, China}
\affiliation[b]{Kavli Institute for Astronomy and Astrophysics, Peking
University, \\ Beijing 100871, China}

\emailAdd{wzhax@pku.edu.cn}
\emailAdd{lshao@pku.edu.cn}
\emailAdd{lxl@pku.edu.cn}

\abstract{Axion is a popular candidate for dark matter particles. Axionic dark matter may form Bose-Einstein condensate and may be gravitationally bound to form axion clumps. Under the presence of electromagnetic waves with frequency $\omega= m_{a}/ 2$, where $m_a$ is the axion mass, a resonant enhancement may occur, causing an instability of the axion clumps. With analytical and numerical approaches, we study the resonant instability of
axionic dark matter clumps with infinite homogeneous mass distribution, as well as distribution with a finite boundary. After taking realistic astrophysical environments into consideration, including gravitational redshift and
plasma effects, we obtain an instability region in the axion {\it density --
clump size} parameter space with given mass and coupling of axions. In
particular, we show that, for axion clumps formed by the QCD axions in
equilibrium, no resonant instability will occur.}

\keywords{axions, dark matter theory, absorption and radiation processes}

\begin{document}
\maketitle
\flushbottom

\section{Introduction}

Since its establishment, the standard model of particle physics has
become one of the most successful theories in physics. However, challenges
still exist in the standard model, such as topics related to the neutrino mass,
the hierarchy problem, the strong charge-parity (CP) problem, and so on. It is thus
important to go beyond the standard model. The strong CP problem asks why
the CP symmetry is preserved in the strong interaction. It is elegantly solved
by introducing the Peccei-Quinn (PQ) symmetry~\cite{ax1,ax2}. At low
energies, the PQ symmetry is spontaneously broken and a pseudo-Goldstone boson,
the quantum chromodynamics (QCD) axion, is left as a trace of higher energy physics.

The QCD axion has become one of the most promising candidates for new physics beyond the standard model. The mass of QCD axion, $m_{a}$, and the PQ symmetry
breaking scale, $f_{a}$, are related by~\cite{precise,marsh}
\begin{equation}\label{eq:massfa}
m_{a}=5.7\times10^{-6} \, {\rm eV} \left(\frac{10^{12} \, {\rm GeV} }{f_{a}/C} \right) \, ,
\end{equation}
where $C$ is the color anomaly or the domain wall number, and we use $\hbar=c=1$ throughout the paper. In string
theory~\cite{string}, scalar or pseudo-scalar particles naturally arise
from compactification of extra dimensions. These particles do not
necessarily solve the strong CP problem, but behave like the QCD axion in many
ways. They are called axion-like particles (ALPs). For ALPs, the mass,
$m_{a}$, and energy scale, $f_{a}$, are treated as independent
parameters. String theory may imply the existence of many sorts of ALPs called the ``string axiverse''\cite{stringaxiverse,axiverse}.

Axions interact with photons, electrons, and nucleons, making it possible to
detect them in experiments. Axion-photon interaction is the most general
feature of ALPs, which is also the primary channel in axion search
experiments. Ground-based experiments include light shining through
wall~\cite{light}, microwave cavity experiments (e.g. the ADMX
experiment~\cite{ADMX}), and helioscopes (e.g. the CAST
experiment~\cite{CAST}). Axions or ALPs are also expected to be produced in many astrophysical processes, which may have observable effects. So
astrophysical processes can be used to constrain axion parameters. Examples
include the supernova SN\,1987A observations~\cite{SN}, black hole
superradiance~\cite{Superradiance}, and pulsars~\cite{Caputo:2019tms}. In recent years some anomalous observational results, like the stellar cooling excess~\cite{cooling}, 21-cm line observations~\cite{21cm}, and the TeV transparency~\cite{TeV}, are found and are proposed to be probably caused by ALPs.

Axions result from the PQ symmetry breaking. The PQ symmetry breaking scale $f_{a}$ is high, so all the couplings between
axions and standard model particles are suppressed by at least a factor ${1}/{f_{a}}$, making axions an excellent dark matter candidate. The main
mechanisms~\cite{mini} to create dark matter axions in the early universe are
the misalignment mechanism and the decay of strings and domain walls.

We must distinguish the cases whether the PQ symmetry is broken or unbroken during the inflation, which give different axionic dark matter density today. If the PQ symmetry is unbroken during inflation, the initial misalignment angle $\theta_{i}$ takes a random value of order unity in each Hubble patch after the symmetry breaking. For the case that the PQ symmetry is broken during inflation, $\theta_{i}$ takes a single value of order unity. In a Friedmann-Lema\^itre-Robertson-Walker spacetime, the zero mode of the dimensionless axion field $\theta(t)$ evolves according to
$\ddot{\theta}+3H(t)\dot{\theta}+m_{a}^{2}(t)\theta=0$, where $H(t)$ is the
Hubble parameter. When $H(t)\sim m_{a}(t)\sim 1/t$, the axion field started
to oscillate. The axion mass depends on the temperature and hence on the
cosmic time $t$. At a high temperature $T \sim {\cal O} \left({\rm
GeV}\right)$, the interacting instanton liquid model
gives~\cite{instantonliquid},
\begin{equation}\label{eq:masstem}
m_{a}(T)=3.07\times10^{-9}\, {\rm eV}\left(\frac{f_{a}}{10^{12}\, {\rm GeV}}\right)^{-1}\left(\frac{T}{{\rm GeV}}\right)^{-3.34} \, .
\end{equation}
We define the time $t_{0}$ when the axion field started to oscillate by $m_{a} t_{0}=1$. The relation between the time $t$ and the temperature $T$ can be determined using $H=1/(2t)$ and $H^{2}=4\pi^{3} GT^{4}\mathcal{N}/45$, where the effective number of particle types $\mathcal{N}\simeq60$ for temperature $T \sim {\cal O} \left({\rm GeV}\right)$~\cite{cosmo}.  Then we use eq.~(\ref{eq:masstem}) to find
$t_{0}=2.71\times10^{-7} \, {\rm s} \left({f_{a}} / {10^{12} \, {\rm GeV}}
\right)^{0.375}$. When the temperature of the Universe drops to the QCD scale, the axion potential becomes periodical due to the instanton effect, and $\theta$ relaxes to the minimum of the potential. After that, the axion mass becomes constant in time and axions behave very much like ordinary matters.

Cosmic strings or domain walls may also form in the early universe. Such
topological defects will decay and emit axions. If the domain wall number
$C=1$, the string-wall system is unstable and quickly decays to axions. But for
$C>1$, the decay of the string-wall system is slow and there is a ``domain wall
problem'' because domain walls may dominate the density of the Universe, in contradiction with cosmological observations. There are some solutions to the domain wall problem~\cite{cosmo}, but for simplicity we focus mainly on the $C=1$ case.

After these dark matter axions are created, they become non-relativistic
due to the expansion of the Universe and finally gravitationally bound. These
dark matter axions are proposed to form clumps or miniclusters in
galaxies~\cite{mini,largemis,clump2}, which are localized regions of high axion density.

The characteristic size of axion miniclusters can be estimated as follows. A characteristic length at $t_{0}$, which is roughly the size of particle horizon, expanded until the matter-radiation equality era, when gravity started to dominate and the size of miniclusters stopped to increase since then. Assuming no late time collapse, the size of the minicluster is then given by~\cite{mini},
\begin{equation}
l_{\rm mc}\sim t_{0}\frac{a_{\rm eq}}{a(t_{0})}\sim\sqrt{t_{0}t_{\rm
eq}}=2.0\times10^{13} \, {\rm cm}\left(\frac{f_{a}}{10^{12}\, {\rm
GeV}}\right)^{0.187} \, .
\end{equation}
The characteristic size of axion miniclusters is thus about $10^{-5}\, {\rm
ly}$. The characteristic density equals to the density of axions at the
matter-radiation equality era, of order $10^{-19} \, {\rm g/cm^{3}}$. Later
axion miniclusters may collapse and form axion clumps with higher densities.
So it is reasonable to consider those axion clumps with higher densities.

Although axion self-interaction is very weak, in axion miniclusters or
clumps, where the density of axions is relatively high compared with the
background, thermal equilibrium is possible and the whole clumps can be
viewed as a Bose-Einstein condensate (BEC)~\cite{BEC} with a large
occupancy number. In such a high occupancy case, classical particle
description (like in the Boltzmann equation) is not a good approximation,
instead the system is well described by a classical field~\cite{correlation}.

An interesting phenomenon is the stimulated decay of axions. The
spontaneous decay rate of axions is very small. However, in the presence
of an electromagnetic wave with certain frequency, the stimulated
decay $a\rightarrow\gamma+\gamma$ is possible, whose rate can be very
high. The electromagnetic wave will be greatly enhanced during the
process. Such a parametric resonance process is possible for axions in a dark matter halo~\cite{resonance,stidec}, and for ultralight axions appearing in black hole superradiance~\cite{stidecay}. Parametric resonance in the presence of both axionic dark matter and magnetic fields is studied in ref.~\cite{sta}. Resonance should also be possible in axion clumps~\cite{axlaser}. Parametric resonance in axion clumps is studied in
ref.~\cite{resonance2}, where the growth rate has been calculated. By studying the resonant enhancement in an axion condensate, whose size is as large as a galaxy, severe constraints have been put in the
axion parameter space~\cite{reson}.

We first review the axion-photon interaction~\cite{marsh} in
section~\ref{sec:axion:photon}. Then in section~\ref{subsec:resonance1} we introduce the resonant instability in a homogeneous axion field, which has been discussed in refs.~\cite{resonance,resonance2}. In section~\ref{subsec:resonance2} we discuss the case of axion clumps and emphasize the difference from the homogeneous case. Resonant conditions are derived analytically and numerically by solving differential equations. The results are consistent with the estimate in refs.~\cite{axlaser,resonance2}. We show the evolution of electromagnetic field and axion field through numerical calculation in section~\ref{subsec:resonance3}. In our work, we take the variation of axion density into consideration, which is usually ignored in the literature. In section~\ref{sec:instability}, we obtain the instability region of axion clumps in the parameter space, where no requirements of gravitational equilibrium are assumed. We consider the effects of plasma, gravitational redshift, self-interaction and velocity. Due to these effects, especially the gravitational redshift, the severe constraints on axion parameters in ref.~\cite{reson} should be relaxed. Finally, when the gravitational equilibrium is assumed, results consistent with ref.~\cite{resonance2} are obtained. Some discussions are
present in the last section.

\section{Axion-photon interaction}
\label{sec:axion:photon}

The most well studied axion interaction is its interaction with photons,
which is the most general property for axions and ALPs. Other interactions
have larger theoretical uncertainties and may be further suppressed. The
Lagrangian density for axions and electromagnetic fields can be written as
\begin{equation}\label{eq:lagrangian}
\mathcal{L}=\frac{1}{2}\partial_{\mu}a\partial^{\mu}a-V(a)-\frac{1}{4}F_{\mu\nu}F^{\mu\nu}-\frac{1}{4}g_{a\gamma}aF_{\mu\nu}\tilde{F}^{\mu\nu} \, ,
\end{equation}
where $g_{a\gamma}$ is the axion-photon coupling and $a$ is the axion field. The axion potential can be written approximately as a cosine potential,
\begin{equation}\label{eq:cosine:potential}
V(a)=\frac{m_{a}^{2}f_{a}^{2}}{C^{2}} \left[1- \cos \left(\frac{Ca}{f_{a}} \right) \right] \, ,
\end{equation}
For a low axion density, self-interaction can be neglected, and eq.~(\ref{eq:cosine:potential}) reduces to,
\begin{equation}
V(a)=\frac{1}{2}m_{a}^{2}a^{2} \, .
\end{equation}
The axion-photon coupling constant is~\cite{marsh},
\begin{equation}
g_{a\gamma}=\frac{\alpha_{\rm EM}}{2\pi (f_{a}/C)}C_{a\gamma} \, ,
\end{equation}
where $C_{a\gamma}$ is a model dependent parameter, usually of order unity.
In the Kim-Shifman-Vainshtein-Zakharov (KSVZ) model~\cite{KSVZ1,KSVZ2},
$C_{a\gamma}=-1.92$ and $C=1$, while in the
Dine-Fischler-Srednicki-Zhitnitsky (DFSZ) model~\cite{DFSZ1,DFSZ2},
$C_{a\gamma}=0.75$ and $C=6$.

Due to the coupling to photons, an axion will decay into two photons. The spontaneous decay rate is~\cite{decay},
\begin{equation}
\Gamma_{a}=\frac{g_{a\gamma}^{2}m_{a}^{3}}{64\pi}=1.1\times10^{-24} \, {\rm s}^{-1} \left(\frac{m_{a}}{\rm eV}\right)^{5} \, .
\end{equation}
If the axion mass $m_{a}\gtrsim 20\, {\rm eV}$, spontaneous decay of axions will lead to excess radiation, which can be used to constrain ALPs~\cite{particle,alpdecay}. The stimulated decay process under electromagnetic waves can be much stronger than the spontaneous decay, which is the main topic of this paper.

To investigate the stimulated decay of axions in axion clumps, we use field equations derived from the Lagrangian (\ref{eq:lagrangian}). For non-relativistic dark matter axions, the spatial gradient term $\nabla a$ can be neglected when compared with ${\partial a}/{\partial t}$. In the Coulomb gauge, the field equations are,
\begin{align}
  \label{eq:fieldeq1} \left(\frac{\partial^{2}}{\partial t^{2}}-\nabla^{2}
  \right)\vec{A} &= -g_{a\gamma}\frac{\partial a}{\partial t}
  \left(\nabla\times \vec{A} \right) \,, \\
  \label{eq:fieldeq2} \left(\frac{\partial^{2}}{\partial
  t^{2}}+m_{a}^{2}\right)a &=g_{a\gamma}\frac{\partial \vec{A}}{\partial
  t}\cdot \left(\nabla\times \vec{A} \right) \,.
\end{align}

Classical field equations are enough to describe the axion BEC, where the
occupancy number is high~\cite{correlation}. Eqs.~(\ref{eq:fieldeq1}) and (\ref{eq:fieldeq2}) are used for the following studies. Note
that the equation for the vector potential differs from the usual Maxwell
equation on the right-hand side. The difference is small due to the
smallness of $g_{a\gamma}$, which is usually ranging from $10^{-15}\, {\rm
GeV}^{-1}$ to $10^{-11} \, {\rm GeV}^{-1}$. The results of CAST gave an
upper limit on the axion-photon coupling, $g_{a\gamma}<0.66\times10^{-10}
\, {\rm GeV^{-1}}$ at $95\%$ confidence level, for axion mass $m_{a}<0.02
\, {\rm eV}$~\cite{CAST2}.

\section{Resonance in axion condensate}
\label{sec:resonance}

In this section we review resonance processes in axion condensate. In section~\ref{subsec:resonance1} we consider the case of homogeneous axion field, which has been discussed in refs.~\cite{resonance,resonance2}. In section~\ref{subsec:resonance2} and section~\ref{subsec:resonance3}  axion clumps with finite boundary are considered, where we calculate the process analytically and numerically.

\subsection{The homogeneous case}
\label{subsec:resonance1}
We review the resonant process in a homogeneous axion field first. We consider resonant enhancement with an electromagnetic wave propagating
in a homogeneous axion background field. 
 Such homogeneous axion background
field is unstable due to gravitational attraction and will collapse to form
axion clumps~\cite{correlation}. As an illustration, we use a simple model that the axion background field is uniform and the density is time-independent. Therefore, the background axion field is
\begin{equation}
a=a_{0}\cos(\omega_{0} t+\varphi)\, .
\end{equation}
For non-relativistic cosmic axions with a low density, the frequency $\omega_{0}=m_{a}$.
We now assume that the amplitude $a_{0}$ is independent of time and position.
The Hamiltonian is
\begin{equation}\label{eq:Hamiltonian}
H=\int {\rm d}^{3}x \, \left[\frac{1}{2}\left(\frac{\partial a}{\partial t}\right)^{2}+\frac{1}{2}(\nabla a)^{2}+\frac{1}{2}m_{a}^{2}a^{2}\right]\, .
\end{equation}
Using eq.~(\ref{eq:Hamiltonian}), a relation between the amplitude $a_{0}$ and the axion density $\rho_{a}$ can be derived,
\begin{equation}
\rho_{a}=\frac{1}{2}m_{a}^{2}a_{0}^{2} \, .
\end{equation}
We consider an incoming monochromatic electromagnetic wave with
frequency $\omega>0$. Assume that the electromagnetic field is weak. Then, $a_{0}$ does not vary and the axion field can be
regarded as a background field. Because the presence of
$\nabla\times\vec{A}$ in eqs.~(\ref{eq:fieldeq1}) and (\ref{eq:fieldeq2}),
it is convenient to use the helicity basis
\begin{equation}
\vec{A}=A_{\pm}(t)e^{ikz}\vec{e}_{\pm} \, ,
\end{equation}
where $\vec{e}_{\pm}=\vec{e}_{x}\pm i\vec{e}_{y}$.
Eq.~(\ref{eq:fieldeq1}) gives~\cite{resonance2}
\begin{equation}\label{eq:ma}
\left(\frac{\partial^{2}}{\partial t^{2}}+k^{2}\right)
A_{\pm}(t)=\pm g_{a\gamma}a_{0}\omega_{0}k \sin(\omega_{0}t+\varphi)A_{\pm}(t) \, .
\end{equation}
The Mathieu equation (\ref{eq:ma}) has a resonant solution where
$A_{\pm}(t)$ grows exponentially with time. If the density of background
axions is low, the resonance happens at $\omega=\omega_{0}/2$ with a small
bandwidth, which corresponds to the stimulated decay
process $a\rightarrow\gamma+\gamma$. For an axion clump with high density, processes like $a+a\rightarrow\gamma+\gamma$ are also possible. Here we consider the process $a\rightarrow\gamma+\gamma$ first.

$A_{\pm}(t)$ can be written as,
\begin{equation}
A_{\pm}(t)=f_{\pm T}(t)e^{-i\left(\omega t+\frac{\varphi}{2}\right)}+f_{\pm
R}(t)e^{i\left(\omega t+\frac{\varphi}{2}\right)} \, ,
\end{equation}
where we separate $A_{\pm}(t)$ into slowly varying amplitudes, $f_{\pm
T}(t)$ and $f_{\pm R}(t)$, and fast varying phases. ``$T$'' and ``$R$''
denote electromagnetic waves propagating in $+z$ and $-z$ directions
respectively. Reflection appears if $f_{\pm R}(t)\neq 0$. As just mentioned, we
require $f_{\pm T}(t)$ and $f_{\pm R}(t)$ to be slowly varying, i.e.,
\begin{equation}\label{eq:slowcon1}
\frac{\dot{f}_{\pm T,R}(t)}{f_{\pm T,R}(t)} \ll \omega \, ,
\end{equation}
\begin{equation}\label{eq:slowcon2}
\frac{\ddot{f}_{\pm T,R}(t)}{\dot{f}_{\pm T,R}(t)} \ll \omega \, .
\end{equation}
Dropping all fast varying terms, we get,
\begin{align}\label{eq:slow1}-4\omega\dot{f}_{\pm R}(t)+2i(k^{2}-\omega^{2})f_{\pm R}(t)&=\pm g_{a\gamma}a_{0}\omega_{0}kf_{\pm T}(t)e^{i(\omega_{0}-2\omega)t} \, , \\
\label{eq:slow2}4\omega\dot{f}_{\pm T}(t)+2i(k^{2}-\omega^{2})f_{\pm T}(t)&=\mp g_{a\gamma}a_{0}\omega_{0}kf_{\pm R}(t)e^{i(2\omega-\omega_{0})t} \, .
\end{align}
In vacuum, we take the dispersion relation $\omega=k$ for
electromagnetic waves. Resonance occurs near
$\omega=\omega_{0}/2$. Denote $\omega-\omega_{0}/2=\epsilon$ and define
$f_{\pm T}(t)=e^{i\epsilon t}F_{\pm T}(t)$, $f_{\pm R}(t)=e^{-i\epsilon
t}F_{\pm R}(t)$. From above equations we have
\begin{align}
\dot{F}_{\pm T}(t)+i\epsilon F_{\pm T}(t)&=\mp\frac{1}{2}g_{a\gamma}a_{0}kF_{\pm R}(t) \, , \\
\dot{F}_{\pm R}(t)-i\epsilon F_{\pm R}(t)&=\mp\frac{1}{2}g_{a\gamma}a_{0}kF_{\pm T}(t) \, .
\end{align}
The equations can be solved by taking derivatives on both sides, which leads to
\begin{align}
\ddot{F}_{\pm T}(t)+\left(\epsilon^{2}-\frac{1}{4}g_{a\gamma}^{2}a_{0}^{2}k^{2}\right) F_{\pm T}(t)&=0 \, , \\
\ddot{F}_{\pm R}(t)+\left(\epsilon^{2}-\frac{1}{4}g_{a\gamma}^{2}a_{0}^{2}k^{2}\right) F_{\pm R}(t)&=0 \, .
\end{align}
After taking initial conditions $F_{\pm T}(0)=f_{\pm T}(0)$ and $F_{\pm R}(0)=0$, and defining
$\lambda^{2}=\frac{1}{4}g_{a\gamma}^{2}a_{0}^{2}k^{2}-\epsilon^{2}$, one
has
\begin{align}
\label{eq:solution1}F_{\pm T}(t)&=f_{\pm T}(0)\left[ \cosh(\lambda t)-\frac{i\epsilon}{\lambda} \sinh(\lambda t)\right] \, , \\
\label{eq:solution2}F_{\pm R}(t)&=\mp\frac{g_{a\gamma}a_{0}k}{2\lambda}f_{\pm T}(0) \sinh(\lambda t) \, .
\end{align}
These exponentially growing solutions show that there is a
resonant enhancement if $\lambda^{2}>0$. Note that the
exponential growth can not persist forever, because
the density of the axion field decreases as axions decay to photons. The condition for resonance $\lambda^{2}>0$ yields
\begin{equation}
-\frac{1}{2}g_{a\gamma}\sqrt{\frac{\rho_{a}}{2}}<\epsilon<\frac{1}{2}g_{a\gamma}\sqrt{\frac{\rho_{a}}{2}} \, .
\end{equation}
The resonant bandwidth for homogeneous axion condensate is $\Delta\omega=g_{a\gamma}\sqrt{{\rho_{a}}/{2}}$. The relative bandwidth is
\begin{equation}\label{eq:bandwidth}
\frac{\Delta\omega}{\frac{1}{2}m_{a}}=\frac{2g_{a\gamma}}{m_{a}}\sqrt{\frac{\rho_{a}}{2}}=9.28\times10^{-20}\left(\frac{g_{a\gamma}}{10^{-12}\, {\rm GeV}^{-1}}\right)
\left(\frac{10^{-2}\, {\rm eV}}{m_{a}}\right)\left(\frac{\rho_{a}}{10^{-19} \, {\rm g/cm^{3}}}\right)^{\frac{1}{2}} \, .
\end{equation}
It is clear that for the low axion background density in axion clumps, the resonant bandwidth is very narrow. As a result, the resonance is easily destroyed (see section~\ref{sec:instability}). The maximum Floquet exponent, namely the maximum exponential growth rate, is ${g_{a\gamma}}\sqrt{{\rho_{a}}/{8}}$.

The resonance process $a+a\rightarrow\gamma+\gamma$ is also possible. It corresponds to a resonance at electromagnetic wave frequency $\omega=\omega_{0}$. In this case a different ansatz of $A_{\pm}(t)$ is needed,
\begin{equation}\label{eq:ansatz}
A_{\pm}(t)=f_{\pm T}(t)e^{-i(\omega t+\varphi)}+f_{0}(t)+f_{\pm R}(t)e^{i(\omega t+\varphi)} \, ,
\end{equation}
where $f_{\pm T}(t)$, $f_{0}(t)$, $f_{\pm R}(t)$ are slowly varying
functions compared with the time scale $1/\omega$. Consequently, we can drop
all second derivatives with respect to $t$. A zero-frequency term
$f_{0}(t)$ has been added in eq.~(\ref{eq:ansatz}). After dropping fast varying terms, we have
\begin{align}
    \dot{f}_{\pm,T}&=\mp \frac{Bf_{0}}{2\omega}e^{i\epsilon t}   \,, \\
  f_{0}&=\mp\frac{iB}{k^{2}}\left(f_{\pm,T}e^{-i\epsilon t}-f_{\pm,R}e^{i\epsilon t}\right) \, , \\
  \dot{f}_{\pm,R}&=\mp \frac{Bf_{0}}{2\omega}e^{-i\epsilon t}   \, ,
\end{align}
where we have defined $B=\frac{1}{2}g_{a\gamma}a_{0}\omega_{0}k$ and
$\epsilon=\omega-\omega_{0}$, and we have used eq.~(\ref{eq:ma}). We
eliminate $f_{0}$ in above equations, and further define $f_{\pm
T}(t)=e^{i\epsilon t}F_{\pm T}(t)$ and $f_{\pm R}(t)=e^{-i\epsilon t}F_{\pm
R}(t)$. Then, we have
\begin{align}
    \dot{F}_{\pm,T}+i\epsilon F_{\pm,T}&= \frac{iB^{2}}{2\omega^{3}}\left(F_{\pm,T}-F_{\pm,R}\right)   \,, \\
  \dot{F}_{\pm,R}-i\epsilon F_{\pm,R}&= \frac{iB^{2}}{2\omega^{3}}\left(F_{\pm,T}-F_{\pm,R}\right)   \, .
\end{align}
After taking time derivative of the first equation, a second-order differential equation for $F_{\pm,T}$ is obtained,
\begin{equation}
\ddot{F}_{\pm,T}=\left[\frac{B^{4}}{4\omega^{6}}-\left(\frac{B^{2}}{2\omega^{3}}-\epsilon\right)^{2}\right]F_{\pm,T} \, .
\end{equation}
Exponentially growing solutions exist with a bandwidth
$\Delta\omega=B^{2}/\omega^{3}$. The maximum Floquet exponent is
${g_{a\gamma}^{2}\rho_{a}}/{4m_{a}}$, which is small compared with the maximum Floquet exponent of $\omega=\omega_{0}/2$, which is ${g_{a\gamma}}\sqrt{{\rho_{a}}/{8}}$.
The resonant bandwidth is also small, making the exponential growth almost
impossible in realistic axion clumps. Therefore the high-order process $a+a\rightarrow\gamma+\gamma$ can always be neglected, at
least for axion clumps that are not extremely dense.

\subsection{The effects of finite size}
\label{subsec:resonance2}
We now turn to the finite size case, namely an axion clump. In general, an axion clump refers to dark matter axions that are gravitationally bounded in a localized region. Axion miniclusters
may be composed of axion clumps and possibly some diffuse axions. As we will see, resonance occur in an axion clump only if its density and size are large enough. This is because axion clumps have boundaries, those photons that reach the boundary will escape. If an axion clump is small, photons will escape before the exponential growth can occur. Resonant enhancement requires that the rate of photon creation by stimulated decay of axions is larger than the rate of photon escape.

Alternatively, we can understand the effect of finite size from the
solutions in eqs.~(\ref{eq:solution1}) and (\ref{eq:solution2}) for a homogeneous axion condensate. We note that initially only the transmitted wave is present, later the reflected wave appears. If the size of the axion
condensate is small, the reflected wave is very weak and can be neglected,
in which case the transmitted wave only becomes a little bit stronger and
the enhancement is usually hard to observe due to the narrow enhancement
bandwidth. However, if the reflected wave is stronger than the initially
transmitted wave before it reaches the boundary of axion condensate,
multiple reflections will be important and the electromagnetic wave will
continue to grow.

We see from the results in eqs.~(\ref{eq:solution1}) and (\ref{eq:solution2}) that in the homogeneous case the reflected wave is stronger than the
initially transmitted wave if
$g_{a\gamma}a_{0}k\sinh(\lambda t)/(2\lambda)>1$. Assume that the effective size of an axion clump is $d_{\rm eff}$, and we take $t=d_{\rm eff}$. If we do not take other factors like gravitational redshift into account, the effective size of  an axion clump is equal to its diameter $d$. In section~\ref{sec:instability} we will see cases where $d_{\rm eff}<d$. Resonance requires at least $\lambda_{0} d_{\rm eff}>0.88$, where
$\lambda_{0}=g_{a\gamma}a_{0}k/2=g_{a\gamma}\sqrt{\rho_{a}/8}$. If an axion
clump satisfies the resonance condition $\lambda_{0} d_{\rm
eff}>0.88$, axions will decay via stimulated emission until the
resonance condition is destroyed due to the decrease of density, whereafter a strong monochromatic electromagnetic wave will be released. We will see later that the resonance condition $\lambda_{0} d_{\rm eff}>0.88$ agrees with numerical results except a factor of order unity. The result of resonance condition estimated here is consistent with the previous estimate $\lambda_{0} d_{\rm eff}>1$ in refs.~\cite{resonance2,axFRB}.

The analysis above gives a physical picture for understanding the effect of boundaries. In the following we will obtain the resonance condition by solving partial differential equations, from which the result becomes more accurate for different axion density distributions. Here we consider two different density distributions of axion clumps: one is a uniform density within a radius $R$, and the other is the equilibrium density distribution. Let us assume a $1+1$ dimensional model (1 space dimension and 1 time dimension), where the electromagnetic field propagates in the $z$ direction and the axion clump density depends on $z$. It is a good approximation if the electromagnetic wave is a plane wave and the $x,y$ dependence of the density is weak. In fact, we find that the $x$ and $y$ dependence only contributes higher order terms in field equations, as long as the spatial variation length is larger than the wavelength of plane waves. So the system effectively has one spatial dimension. The amplitudes of axion background field and electromagnetic waves also change with time, as
\begin{align}
\vec{A}(z,t)=&\frac{1}{2}f_{+T}(z,t)e^{i(kz-\omega
t-\frac{\varphi}{2})}\vec{e}_{+}+\frac{1}{2}f_{+R}(z,t)e^{i(kz+\omega
t+\frac{\varphi}{2})}\vec{e}_{+} \nonumber \\
&+\frac{1}{2}f_{-T}(z,t)e^{i(kz-\omega
t-\frac{\varphi}{2})}\vec{e}_{-}+\frac{1}{2}f_{-R}(z,t)e^{i(kz+\omega
t+\frac{\varphi}{2})}\vec{e}_{-}+ {\rm c.c.} \, , \\
a(z,t)=&\frac{1}{2}\alpha(z,t)e^{-i(\omega_{0}t+\varphi)}+ {\rm c.c.} \, .
\end{align}
The electromagnetic waves have two polarization modes, ``$+$'' and ``$-$'', and
propagate in $+z$ and $-z$ directions for ``$T$'' and ``$R$'', respectively.
Compared to phases,
$f_{\pm T}(z,t)$, $f_{\pm R}(z,t)$ and $\alpha(z,t)$ are slowly varying functions with respect to $z$ and $t$. The factor $1/2$ is introduced so that the ansatz is in accord with previous discussions in the homogeneous case. We take $\omega_{0}=m_{a}$, $k=\omega$ and
$\epsilon=\omega-\omega_{0}/2$. Dropping all higher-order terms
and fast varying terms yields,
\begin{align}
\label{eq:a0}\frac{\partial\alpha}{\partial
t}&=\frac{m_{a}}{4}g_{a\gamma}\left(f^{*}_{+R}f_{+T}-f^{*}_{-R}f_{-T}\right)e^{-2i\epsilon
t} \, , \\
\label{eq:b0}\frac{\partial f_{\pm T}}{\partial t}+\frac{\partial f_{\pm
T}}{\partial z}&=\mp\frac{1}{4}g_{a\gamma}m_{a}\alpha f_{\pm
R}e^{2i\epsilon t} \, , \\
\label{eq:c0}\frac{\partial f_{\pm R}}{\partial t}-\frac{\partial f_{\pm
R}}{\partial z}&=\mp\frac{1}{4}g_{a\gamma}m_{a}\alpha^{*} f_{\pm
T}e^{-2i\epsilon t} \, .
\end{align}
If we take the incoming electromagnetic wave frequency to be $\omega=m_{a}/2$, all the amplitudes can be taken as real numbers,
\begin{align}
  \label{eq:a} \frac{\partial\alpha}{\partial t}
  &= \frac{m_{a}}{4}g_{a\gamma}\left(f_{+R}f_{+T}-f_{-R}f_{-T}\right) \, ,
  \\
  \label{eq:b} \frac{\partial f_{\pm T}}{\partial t}+\frac{\partial f_{\pm
  T}}{\partial z}&=\mp\frac{1}{4}g_{a\gamma}m_{a}\alpha f_{\pm R} \, , \\
  \label{eq:c} \frac{\partial f_{\pm R}}{\partial t}-\frac{\partial f_{\pm
  R}}{\partial z}&=\mp\frac{1}{4}g_{a\gamma}m_{a}\alpha f_{\pm T} \, .
\end{align}

As the first example, we assume that the amplitude $\alpha(z,t)$ is a real constant $a_{0}$ in the region $0<z<2R$, where $R$ is the radius of the axion clump. In $z>2R$ or $z<0$ the density of axions is zero. The boundaries at $z=0$ and $z=2R$ are assumed to be smooth enough so that gradient terms are not important. We have assumed that the axion density does not change with time, which is true for weak electromagnetic fields. We consider the ``$+$'' mode, that is, we take $f_{-T}(z,t)=f_{-R}(z,t)=0$. For the ``$-$'' mode, the results are similar. If both ``$+$'' mode and ``$-$'' mode are present, the overall physical picture will be the same. Initially, only a transmitted wave propagating in $+z$ direction with a small amplitude $F_{0}$ is present. We also note that there is no reflected wave at $z=2R$. Hence the boundary and initial conditions are,
\begin{align}
\label{eq:d}f_{+T}(0,t)&=F_{0} \, , \\
\label{eq:e}f_{+R}(2R,t)&=0 \, , \\
\label{eq:f}f_{+T}(z,0)&=F_{0} \, , \\
\label{eq:g}f_{+R}(z,0)&=0 \, .
\end{align}
In numerical calculations, we find that, for an axion clump with a uniform density, the equations with above boundary conditions give exponentially growing solutions if
\begin{equation}\label{eq:resonantcon}
g_{a\gamma}\sqrt{\frac{\rho_{a}}{8}}d_{\rm eff} \gtrsim \frac{\pi}{2} \, .
\end{equation}
It can be numerically calculated that the factor on the right-hand side of eq.~\eqref{eq:resonantcon} is between 1.570 to 1.572, which we tentatively guess  to be $\pi/2$. Numerical
solution shows that if $g_{a\gamma}d_{\rm eff}\sqrt{\rho_{a}/8}<\pi/2$, the
amplitude of electromagnetic wave will stop increasing and become constant
in time, in which case almost no observable effects occur.

A heuristic analysis may show the origin of the factor $\pi/2$. We search for the critical point where the solution $f_{+T}$ does not become constant in time as $t\rightarrow+\infty$, which can be interpreted as the critical point for the exponential growth. Eqs.~(\ref{eq:b}) and (\ref{eq:c}) can be combined to give a single equation,
\begin{equation}
\frac{\partial^{2} f_{+T}}{\partial t^{2}}-\frac{\partial^{2} f_{+T}}{\partial z^{2}}-\frac{1}{16}g_{a\gamma}^{2}m_{a}^{2}a_{0}^{2} f_{+ T}=0 \, .
\end{equation}
The boundary conditions in eqs.~(\ref{eq:d}) and (\ref{eq:e}) can be rewritten.
We define $f_{+T}(z,t)=F_{0}+F_{+T}(z,t)$ and then use eq.~(\ref{eq:b}). The
boundary conditions become $F_{+T}(0,t)=0$ and $\left(\partial
F_{+T}/\partial t+\partial F_{+T}/\partial z\right)|_{z=2R}=0$. As $t\rightarrow+\infty$, $\partial F_{+T}/\partial t=0$; so in the limit we can regard the boundary conditions as $F_{+T}(0,t)=0$ and $\partial F_{+T}/\partial z|_{z=2R}=0$. Thus we separate the variables as usual for $t\rightarrow+\infty$,
\begin{equation}
F_{+T}=\sum\limits_{n}T_{n}(t)\sin\frac{\left(n+\frac{1}{2}\right)\pi z}{2R} \, ,
\end{equation}
where, after dropping the incoming electromagnetic wave amplitude
$F_{0}$, which is negligibly small, $T_{n}(t)$ satisfies
\begin{equation}
T_{n}''(t)+\left[\frac{(n+\frac{1}{2})^{2}\pi^{2}}{4R^{2}}-\frac{1}{16}g_{a\gamma}^{2}m_{a}^{2}a_{0}^{2}\right]T_{n}(t)=0 \, .
\end{equation}
For the result to be consistent as $t\rightarrow+\infty$, we require that
$T_{n}(t)$ should not exponentially grow. For the $n=0$
mode, which is the most likely to grow exponentially, we require
$\pi^2/(16R^2)-g_{a\gamma}^2m_{a}^2a_{0}^2/16<0$. Using
$\rho_{a}=m_{a}^{2}a_{0}^{2}/2$ for a uniform axion clump of radius $R$ and
define $d_{\rm eff}=2R$, we find the resonance condition to be
$g_{a\gamma}d_{\rm eff}\sqrt{\rho_{a}/8}>\pi/2$. It can be
written as,
\begin{equation}
\rho_{a}>1.99\times10^{-21}\, {\rm g/cm^{3}}
\left(\frac{g_{a\gamma}}{10^{-12}\, {\rm
GeV^{-1}}}\right)^{-2}\left(\frac{d_{\rm eff}}{\rm ly}\right)^{-2} \,.
\end{equation}
We further note that the resonance condition $g_{a\gamma}d_{\rm
eff}\sqrt{\rho_{a}/8}>\pi/2$ is the same up to a factor of order unity as the
condition for the monochromatic approximation. Because the finite
size of an axion clump, the wave frequency will have a characteristic
bandwidth of order $1/d$, where $d$ is the size of the axion clump. If the
characteristic bandwidth of initial wave is smaller than the
resonant bandwidth in eq.~(\ref{eq:bandwidth}) in the homogeneous case, resonance will occur.

If an axion clump is in equilibrium, a profile
$\alpha(z)=\alpha_{0}\sech(z/R)$ is a good approximation~\cite{clump}, where $R$ is characteristic radius of the axion clump. A similar resonance condition for axion clump in equilibrium is,
\begin{equation}\label{eq:resonancecon2}
2R\frac{g_{a\gamma}}{2}\sqrt{\frac{\rho_{a}}{2}}>1.00 \, ,
\end{equation}
where $\rho_{a}$ is the density of axion clump in the center. The result is
obtained in numerical calculations. It is interesting to note that, the resonance condition in eq.~(\ref{eq:resonancecon2}) for axion clumps in equilibrium derived from numerical calculations is almost exactly the same as the estimate $\lambda_{0}d_{\rm eff}>1$ in refs.~\cite{axFRB,resonance2}. For the uniform density case the factor on the right-hand side becomes $\pi/2$.

\subsection{Numerical calculation}
\label{subsec:resonance3}

\begin{figure}[htbp]
  \centering
  \includegraphics[width=15cm]{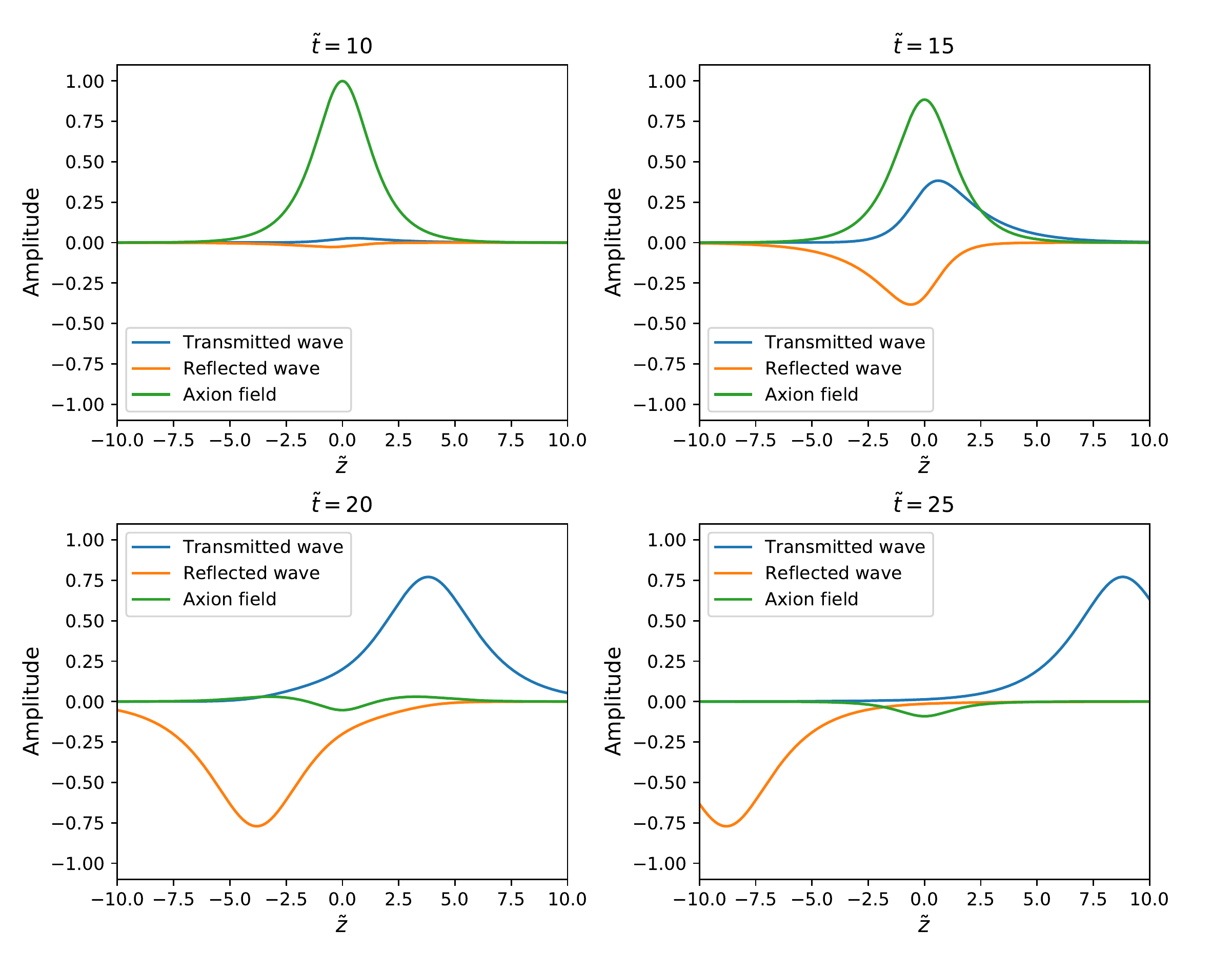}\\
  \caption{Time evolution slices at $\tilde{t}=10$ (upper left), $\tilde{t}=15$ (upper right), $\tilde{t}=20$ (lower left), and $\tilde{t}=25$ (lower right) for the dimensionless amplitudes of electromagnetic waves and the axion field. Initially the amplitude for the axion field is $\tilde{\alpha}(\tilde{z},0)=\sech(\tilde{z}/1.1)$ and the amplitude for the incoming electromagnetic wave is $\tilde{f}_{+T}=0.0001$. As time goes on, the axion clump decays and the two electromagnetic waves propagate in opposite directions. If we take $g_{a\gamma}=10^{-12}\, {\rm GeV^{-1}}$ and
$\rho_{a}=10^{-9}\, {\rm g/cm^{3}}$, the unit time scale is $28.36 \,
{\rm sec}$ and the unit length scale is $9.0\times10^{-7}\, {\rm ly}$.}
  \label{imga}
\end{figure}

\begin{figure}[htbp]
  \centering
  \begin{minipage}[t]{7cm}
  \subfigure[]{\includegraphics[width=7cm]{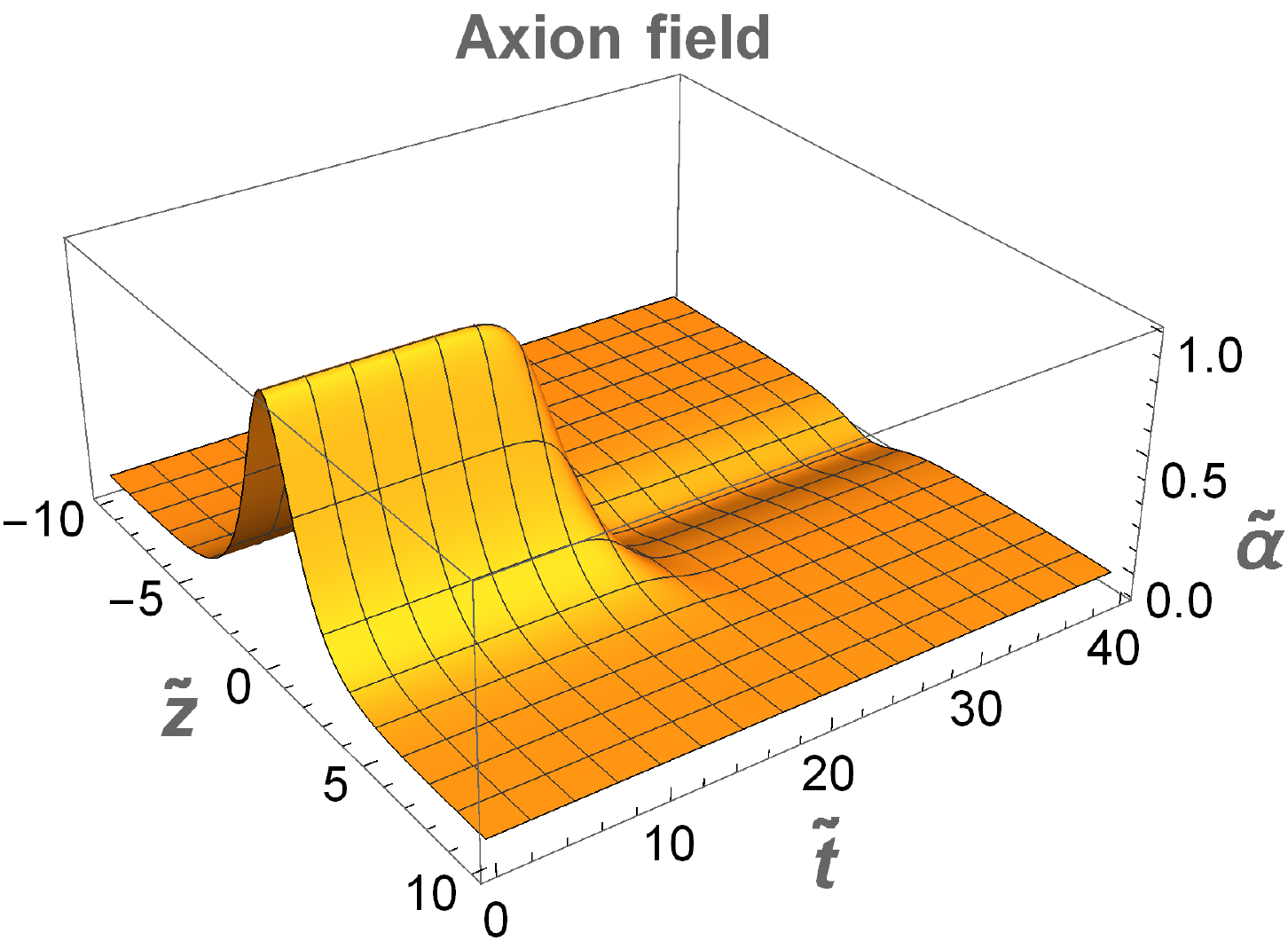}}
  \end{minipage}
  \begin{minipage}[t]{7cm}
  \subfigure[]{\includegraphics[width=7cm]{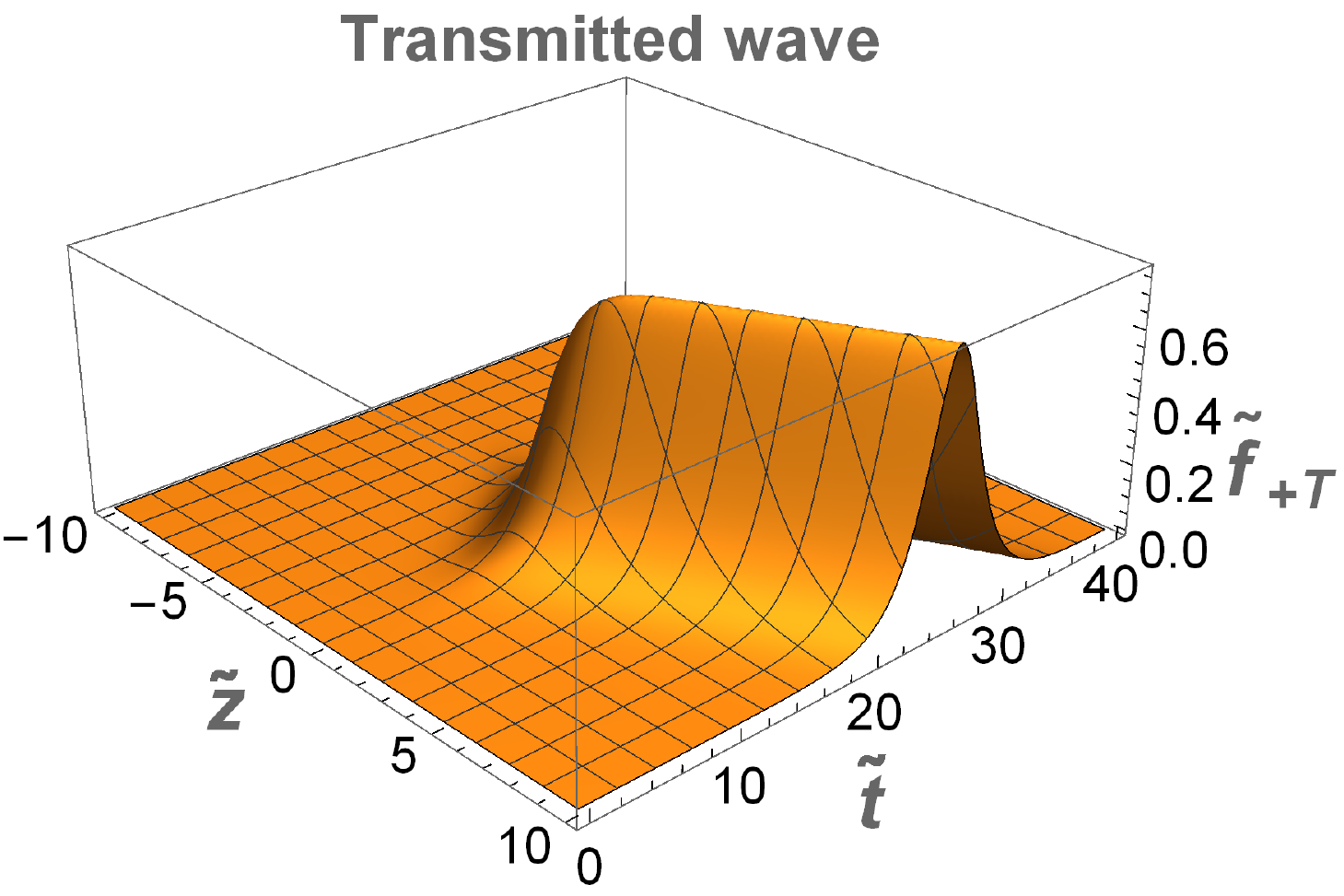}}
  \end{minipage}
  \begin{minipage}[t]{7cm}
  \subfigure[]{\includegraphics[width=7cm]{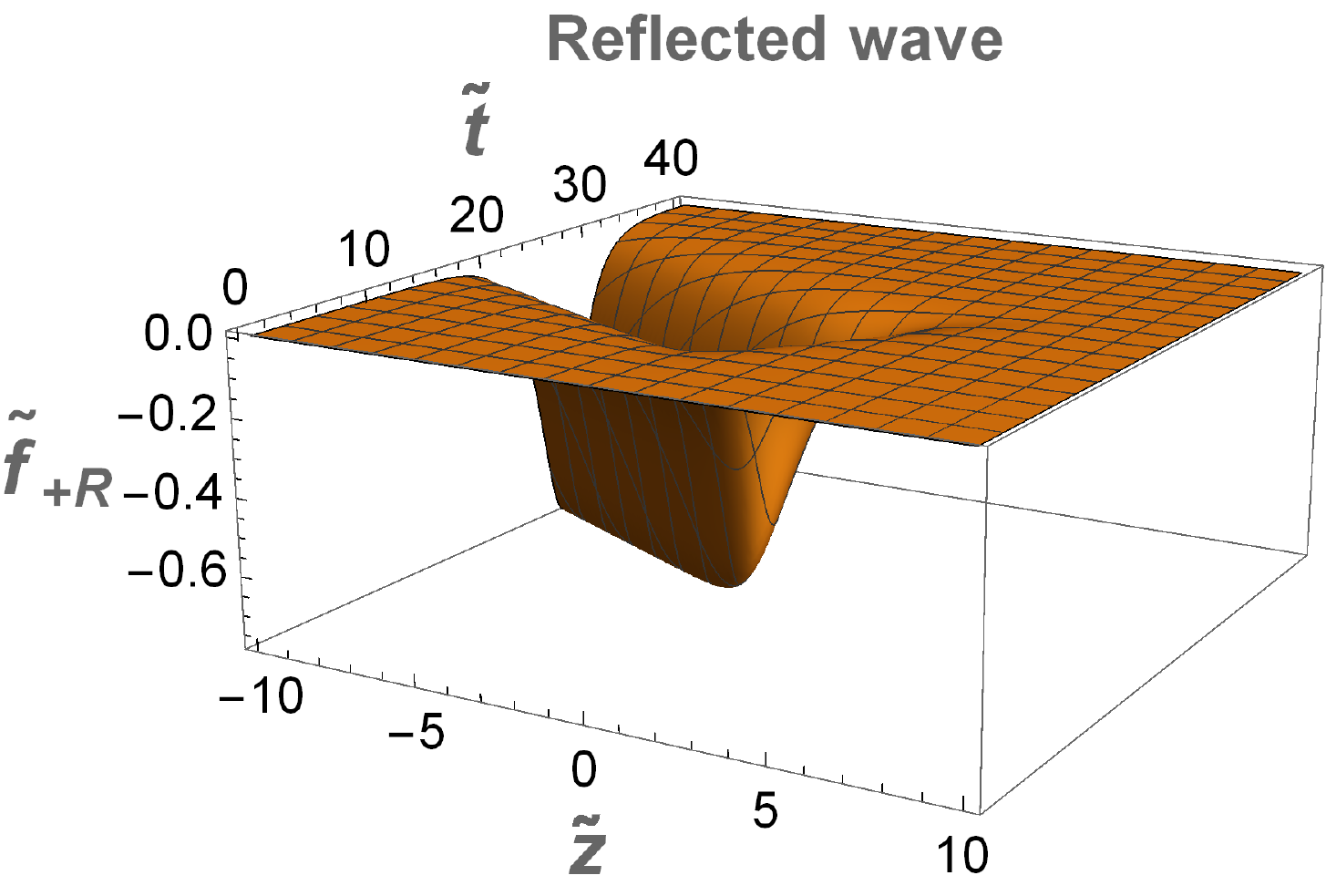}}
  \end{minipage}\\
  \caption{Evolution of the axion field (a), the transmitted wave (b) and the reflected wave (c). At time $\tilde{t}=0$, only a weak incoming electromagnetic wave and a normal axion clump are present. During
the resonance, the electromagnetic wave grows exponentially while
the axion clump density decreases sharply. The height
of the surface represents the amplitude as a function of (rescaled) space $\tilde{z}$ and time $\tilde{t}$. Note that the density
and intensity are proportional to the squared amplitude. Initially
the amplitude of the axion field is
$\tilde{\alpha}(\tilde{z},0)=\sech(\tilde{z}/1.1)$. The axion clump decays, and the two electromagnetic waves propagate in opposite directions. For $\tilde{R}=1.1$ and an initial condition $\tilde{f}_{+T}=0.0001$, the electromagnetic wave becomes strong after $\tilde{t}=10$. For a weaker incoming electromagnetic wave, a longer time is needed for the resonance. If we take $g_{a\gamma}=10^{-12}\, {\rm GeV^{-1}}$ and $\rho_{a}=10^{-9}\, {\rm g/cm^{3}}$, the unit time scale is $28.36 \, {\rm sec}$ and the unit length scale is $9.0\times10^{-7}\, {\rm ly}$.}
  \label{imgb}
\end{figure}

Now we want to calculate the complete process of resonant
enhancement and the evolution of axion clump density. The density of an axion clump is no longer constant in time as was assumed in previous papers~\cite{resonance2,reson}. We use an initial
profile~\cite{clump} $\alpha(z,0)=\alpha_{0}\sech(z/R)$, where $R$
is a characteristic radius of the axion clump. The profile is a good
approximation for axion clumps that have reached equilibrium
configurations~\cite{clump}. Other boundary and initial conditions are similar as in the last subsection. In the numerical calculation, it is useful to use dimensionless variables. Define dimensionless time and position,
\begin{align}
\tilde{t}&=\frac{g_{a\gamma}}{2}\sqrt{\frac{\rho_{a}}{2}}t \, , \\
\tilde{z}&=\frac{g_{a\gamma}}{2}\sqrt{\frac{\rho_{a}}{2}}z \, .
\end{align}
Here $\rho_{a}$ is the density in the center of the axion clump at
$t=0$. In general $\rho_{a}$ is not the average density at $t=0$ unless the axion clump is uniform. We also introduce a dimensionless amplitude of the electromagnetic wave and the axion field,
\begin{equation}
\tilde{f}_{\pm T,R}=\frac{m_{a}}{2}\sqrt{\frac{2}{\rho_{a}}}f_{\pm T,R} \, ,
\end{equation}
\begin{equation}
\tilde{\alpha}=\frac{m_{a}}{2}\sqrt{\frac{2}{\rho_{a}}}\alpha \, .
\end{equation}
We take $f_{-T}(z,t)=f_{-R}(z,t)=0$ so that only ``+'' mode is considered, which has no effect on the physical picture but will simplify calculations. We also take the incoming electromagnetic wave frequency $\omega=m_{a}/2$ so that all the amplitudes can be taken as real numbers. The equations become
\begin{align}
\frac{\partial\tilde{\alpha}}{\partial \tilde{t}}&=\tilde{f}_{+R}\tilde{f}_{+T} \, , \\
\frac{\partial \tilde{f}_{+T}}{\partial \tilde{t}}+\frac{\partial \tilde{f}_{+T}}{\partial \tilde{z}}&=-\tilde{\alpha} \tilde{f}_{+R} \, , \\
\frac{\partial \tilde{f}_{+R}}{\partial \tilde{t}}-\frac{\partial \tilde{f}_{+R}}{\partial \tilde{z}}&=-\tilde{\alpha} \tilde{f}_{+T} \, .
\end{align}
The equations above only apply to the case where the slowly
varying conditions (\ref{eq:slowcon1}) and (\ref{eq:slowcon2}) are satisfied, i.e.,
\begin{equation}
\frac{d\tilde{f}_{+T,R}}{d\tilde{t}}\ll\frac{m_{a}}{g_{a\gamma}}\sqrt{\frac{2}{\rho_{a}}}\tilde{f}_{+T,R} \, ,
\end{equation}
\begin{equation}
\frac{m_{a}}{g_{a\gamma}}\sqrt{\frac{2}{\rho_{a}}}=2.15\times10^{19}\left(\frac{m_{a}}{10^{-2}\, {\rm eV}}\right)\left(\frac{g_{a\gamma}}{10^{-12}\, {\rm GeV^{-1}}}\right)^{-1}
\left(\frac{\rho_{a}}{10^{-19}\, {\rm g/cm^{3}}}\right)^{-\frac{1}{2}} \, .
\end{equation}
The initial axion field is,
\begin{equation}
\tilde{\alpha}(\tilde{z},0)=\sech \left(\frac{\tilde{z}}{\tilde{R}}\right) \, ,
\end{equation}
where $\tilde{R}=g_{a\gamma}R\sqrt{\rho_{a}/8}$ is dimensionless size
of the axion clump. The unit dimensionless length and time scale are,
\begin{equation}
\begin{split}
\frac{1}{g_{a\gamma}\sqrt{\rho_a/8}}
&=2.836\times10^{6} \, {\rm s}\left(\frac{g_{a\gamma}}{10^{-12} \, {\rm GeV}^{-1}}\right)^{-1}\left(\frac{\rho_{a}}{10^{-19}\, {\rm g/cm^{3}}}\right)^{-\frac{1}{2}}\\
&=8.986\times10^{-2}\, {\rm ly} \left(\frac{g_{a\gamma}}{10^{-12}\, {\rm GeV}^{-1}}\right)^{-1}\left(\frac{\rho_{a}}{10^{-19}\, {\rm g/cm^{3}}}\right)^{-\frac{1}{2}} \, .
\end{split}
\end{equation}

In the numerical calculation, we assume
a simplified case and do not take gravity and plasma into account (see section \ref{sec:instability}). Our numerical calculation shows that the resonant instability occurs if $\tilde{R}>0.5$.

As an example, we take $\tilde{R}=1.1$ in the calculation. Our results are given in figures~\ref{imga} and \ref{imgb}. If $g_{a\gamma}=10^{-12} \, {\rm GeV}^{-1}$ and $\rho_{a}=10^{-9} \, {\rm g/cm^{3}}$, then the unit time scale is $28.36 \, {\rm sec}$ and the unit length scale is $9.0\times10^{-7} \, {\rm ly}$. Hence we could convert the dimensionless scales in figures to physical size and time scales. Figures \ref{imga} and \ref{imgb} show the evolution of the axion
clump and electromagnetic wave amplitudes when a weak incoming
electromagnetic wave with frequency $\omega=m_{a}/2$ is present. An initial
incoming wave amplitude $\tilde{f}_{+T}=0.0001$ is used. As we can
see, the axion clump decays, and the electromagnetic waves grow
exponentially until the axion clump density has dropped dissatisfying the resonant
condition. The remaining axion clump density should be lower than the critical density in the resonant condition, but not necessarily the same for different initial axion clump densities. This can easily be verified in numerical calculations. It tells us that if initially there are many axion clumps with different densities, after the resonant decay, the density of axion clumps does not pile up at the critical density of resonance. The remaining axion clump density is still randomly distributed below the critical density for resonance.

\section{Resonant instability in astrophysical background}
\label{sec:instability}

The discussion of resonant enhancement in section~\ref{sec:resonance} is
idealized, while the realistic situation can be much more complicated in  astrophysics. Because the resonant bandwidth is very small, any effects on the frequency of photons or axions may affect or even destroy the resonance. These effects include the plasma effect, the gravitational redshift, the self-interaction of axions, and the motion of axions. Usually these effects lead to the shift of frequency of axions or photons, which would change the resonance if the frequency shift is larger than the bandwidth of the resonance. In this section, we quantitatively consider these effects, and derive the instability region of axion clumps in the parameter space. We do not assume that axion clumps are in equilibrium unless otherwise stated.

\subsection{Plasma effect}

The Universe is filled with plasma and neutral gases. For simplicity we only consider plasma here. Dark matter usually resides in galactic halos, where the typical electron density $n_{e}$ is $0.03 \, {\rm cm^{-3}}$~\cite{resonance2}. Plasma modifies the dispersion relation of photons into $\omega^{2}=k^{2}+\omega_{\rm pl}^{2}$, where $\omega_{\rm pl}$ is the plasma frequency,
\begin{equation}
\omega_{\rm pl}^{2}=\frac{4\pi e^{2}n_{e}}{m_{e}} \, .
\end{equation}
The plasma we considered here is dilute enough so the modification is small. Under the presence of plasma, eqs.~(\ref{eq:slow1}) and (\ref{eq:slow2}) become
\begin{align}
-4\omega\dot{f}_{\pm R}(t)-2i\omega_{\rm pl}^{2}f_{\pm R}(t)&=\pm
g_{a\gamma}a_{0}\omega_{0}kf_{\pm T}(t)e^{i(\omega_{0}-2\omega)t} \, , \\
4\omega\dot{f}_{\pm T}(t)-2i\omega_{\rm pl}^{2}f_{\pm T}(t)&=\mp
g_{a\gamma}a_{0}\omega_{0}kf_{\pm R}(t)e^{i(2\omega-\omega_{0})t} \, .
\end{align}
The difference of $\omega$ and $k$ is at higher orders in the right-hand side of equations, so we use $\omega$ in exchange for $k$ there. If we define $f_{\pm R}(t)=f_{0,\pm R}(t)\exp[-i\omega_{\rm pl}^{2}t/(2\omega)]$ and $f_{\pm T}(t)=f_{0,\pm T}(t)\exp[i\omega_{\rm pl}^{2}t/(2\omega)]$, then $f_{0,\pm R}(t)$ and $f_{0,\pm T}(t)$ satisfy
\begin{align}
-4\omega\dot{f}_{0,\pm R}(t)&=\pm g_{a\gamma}a_{0}\omega_{0}\omega f_{0,\pm
T}(t) \exp \left[ i\left(\omega_{0}-2\omega+\frac{\omega_{\rm
pl}^{2}}{\omega}\right)t \right] \, , \\
4\omega\dot{f}_{0,\pm T}(t)&=\mp g_{a\gamma}a_{0}\omega_{0}\omega f_{0,\pm
R}(t)\exp \left[ i\left(2\omega-\omega_{0}-\frac{\omega_{\rm
pl}^{2}}{\omega}\right)t\right] \, .
\end{align}

These equations are the same as eqs.~(\ref{eq:slow1}) and
(\ref{eq:slow2}) with $\omega=k$, but with frequency shifted as
$\omega\rightarrow \omega-\omega_{\rm pl}^{2}/(2\omega)$. So the plasma effect
shifts the central resonant frequency to
$\omega=\frac{1}{2}m_{a}+\omega_{\rm pl}^{2}/m_{a}$. The relative shift is
$\Delta\omega/\omega=2\omega_{\rm pl}^{2}/m_{a}^{2}$.

The density of plasma is not uniform so in different areas we
have different resonant frequencies. Resonant enhancement
occurs if the relative shift of frequency is smaller than
the relative bandwidth in homogeneous case for a monochromatic electromagnetic wave, namely eq.~(\ref{eq:bandwidth}),
\begin{equation}\label{eq:pla}
\frac{2\omega_{\rm
pl}^{2}}{m_{a}^{2}}<\frac{2g_{a\gamma}}{m_{a}}\sqrt{\frac{\rho_{a}}{2}} \,
,
\end{equation}
that is,
\begin{equation}\label{eq:plasma}
\rho_{a}>7.94\times10^{-18}\, {\rm g/cm^{3}}\left(\frac{g_{a\gamma}}{10^{-12}\, {\rm GeV^{-1}}}\right)^{-2}\left(\frac{m_{a}}{10^{-2}\, {\rm eV}}\right)^{-2}\left(\frac{n_{e}}{0.03\, {\rm cm^{-3}}}\right)^{2} \, .
\end{equation}
The plasma effect is not necessarily negligible as in ref.~\cite{resonance2} for general axion parameters and clump densities discussed here. We stressed that if the plasma has a uniform density, the resonance will
not be stopped because in astrophysical background, the incoming wave has
a continuous spectrum, and the new resonance occurs at the shifted frequency $\omega=\frac{1}{2}m_{a}+\omega_{\rm pl}^{2}/m_{a}$. A non-uniform plasma, with density fluctuations of the same order of $n_{e}$, will lead to changes of the resonant frequency randomly and hence stop the resonance in axion clumps. The discussion here also applies to the case that the incoming electromagnetic wave has a continuous spectrum.

Eq.~(\ref{eq:plasma}) should be regarded as an order of magnitude estimate because of the uncertainty in the density fluctuation of the plasma and the non-uniform density of axion clumps. Besides, for an axion clump whose size is just a little bit larger than what is required by resonance condition (\ref{eq:resonancecon2}), resonance is more sensitive to plasma effect than expected by eq.~(\ref{eq:pla}). So for such axion clumps, the required density for resonance should be larger than eq.~(\ref{eq:plasma}). This is verified in numerical calculations. For axion clumps that are much larger than the size required by resonance condition, however, the required density for resonance may possibly be lower than eq.~(\ref{eq:plasma}). But these effects are hard to deal with quantitatively because the density fluctuations of the plasma are unknown.

\subsection{Gravitational redshift}
\label{subsec:gravity}

The gravity of the axion clump itself and the galaxy it resides leads
to gravitational redshift, which changes the photon frequency but does not
change the frequency of axion field measured in the local observer's proper
frame. Thus gravitational redshift can stop the resonance if the shift of photon frequency is larger than the resonant bandwidth. In the following we consider a finite size axion clump with a uniform density.

The gravitational redshift is,
\begin{equation}
\frac{f_{r}}{f_{e}}=\frac{ \left. \sqrt{-g_{00}}\right|_{e}}{\left.\sqrt{-g_{00}}\right|_{r}} \, ,
\end{equation}
where $f_{e}$ and $f_{r}$ are the frequencies of the photon emitted and received. In the Newtonian limit, which is sufficient here, $-g_{00}=1+2\phi$, where $\phi$ is the gravitational potential. The gravitational redshift in the Newtonian limit can be written as,
\begin{align}
\frac{f_{r}}{f_{e}}=&1+\phi_{e}-\phi_{r} \, , \\
\frac{\Delta f}{f_{e}}=&\phi_{e}-\phi_{r}=-\Delta\phi \, ,
\end{align}
where $\Delta f \equiv f_r - f_e$, $\phi_{e}$ and $\phi_{r}$ are gravitational potentials at the places of photon emitted and received.

We first consider the gravity of a galaxy. Dark matter usually
distributes in galactic halos. The axion clump in
the galactic halo will be affected by the gravitational potential of
a galaxy. Because galaxies have a large variety, only an estimation can be obtained. The incoming electromagnetic waves come from all directions. If resonance occurs in any one direction, the axion clump will be resonantly unstable. So we only need to focuse on the direction that is minimally affected by the gravitational redshift. Photons that propagate almost parallel to the potential isosurfaces will be minimally affected, but the effect is not zero because potential isosurfaces are not planes. As an estimation, we take the radius $R_{0}$ of the potential isosurfaces as the size of galaxies. The size of the axion clump $d$ is assumed to be much smaller than the galactic scale $R_{0}$. We take the gradient of the gravitational
potential as the value near the Solar orbit. The requirement, that the
gravitational redshift does not affect the resonance for incoming
electromagnetic waves almost parallel to the potential isosurface, is
\begin{equation}
\Delta\phi<\frac{\Delta\omega}{\frac{1}{2}m_{a}}=\frac{2g_{a\gamma}}{m_{a}}\sqrt{\frac{\rho_{a}}{2}}
\, .
\end{equation}
Here we use geometrical optics approximation. When an electromagnetic wave propagating through an axion clump, its distance to the center of the galaxy changes slightly, and the maximum change is denoted as $\Delta r$. For incoming
electromagnetic waves almost parallel to the potential isosurface, a simple geometric analysis gives $\Delta r=d^{2}/(8R_{0})$ for $d\ll R_{0}$. The potential difference is $|\Delta\phi|=|\nabla\phi|_{\rm Sun}\Delta r$. Hence we obtain,
\begin{equation}\label{eq:grav1}
\rho_{a}>7.2\times10^{-13}\, {\rm
g/cm^{3}}\left(\frac{g_{a\gamma}}{10^{-12}\, {\rm
GeV^{-1}}}\right)^{-2}\left(\frac{m_{a}}{10^{-2}\, {\rm
eV}}\right)^{2}\left(\frac{R_{0}}{10^{4}\, {\rm
ly}}\right)^{-2}\left(\frac{d}{{\rm ly}}\right)^{4} \, ,
\end{equation}
where $d$ is the size of the axion clump and $R_{0}$ is the radius of the potential isosurfaces. We stress that eq.~(\ref{eq:grav1}) is a rough estimate, which depends on the detail of the galaxy under consideration. In particular, for an axion density of order $10^{-25}\, \rm{g/cm^{3}}$, which is roughly the local dark matter density, gravitational redshift will stop  resonance at the galactic scale, as discussed in ref.~\cite{reson}.

We now consider the gravitational redshift caused by the axion
clump itself. Assume that the axion clump has a uniform density $\rho_{a}$.
The gravitational potential difference between the center and the surface
of an axion clump is $\frac{2}{3}\pi GR^{2}\rho_{a}$. The requirement for the
resonance to occur within the clump radius $R$ is therefore,
\begin{equation}
\frac{\Delta\omega}{\omega}=\frac{2g_{a\gamma}}{m_{a}}\sqrt{\frac{\rho_{a}}{2}}>\frac{2}{3}\pi GR^{2}\rho_{a} \, ,
\end{equation}
\begin{equation}\label{eq:grav2}
\rho_{a}<7.12\times10^{-35}\, {\rm g/cm^{3}}\left(\frac{g_{a\gamma}}{10^{-12}\, {\rm GeV^{-1}}}\right)^{2}\left(\frac{m_{a}}{10^{-2}\, {\rm eV}}\right)^{-2}\left(\frac{d}{\rm ly}\right)^{-4} \, .
\end{equation}

It is worthy to stress the following point. The two constraints in eqs. (\ref{eq:grav1}) and (\ref{eq:grav2}) should not be regarded as requirements for the resonant instability of axion clumps. It is possible that
the resonant enhancement occurs in some regions with sizes $d_{\rm eff}<d$ of
the axion clump, which is not necessarily the whole clump as the two
conditions indicate. So the two constraints are too strong. The $d_{\rm eff}$ is determined by $\Delta\phi(d_{\rm eff})=\sqrt{2\rho_{a}}g_{a\gamma}/m_{a}$, which is the characteristic size that the gravitational redshift does not shift the frequency out of the bandwidth (\ref{eq:bandwidth}).

As an example, we consider an axion clump that satisfies the conditions in eqs.~(\ref{eq:grav1}) and (\ref{eq:grav2}), then the resonant instability
occurs. However, if we increase the size of the axion clump
while keeping the density unchanged, the resonant instability still occurs but
the two conditions may not be met. Actually, the two conditions only give critical densities beyond which the resonant instability occurs.

In real astrophysical situation, the incoming electromagnetic waves have a continuous spectrum. Gravitational redshift leads to a shift in the photon frequency. Originally the resonance occurs near $\omega=m_{a}/2$ with a small bandwidth, then due to the gravitational redshift, the enhanced frequency will be shifted out of the resonant bandwidth. Although there are still some frequencies for the resonance to occur, these frequencies will be shifted out of the bandwidth before resonant enhancement becomes obvious if eq.~(\ref{eq:resonantcon}) is not satisfied. In other words, the gravitational redshift can stop the resonant instability even if the incoming waves have a continuous spectrum.

\subsection{Self-interaction}

Axion self-interaction introduces a correction to the axion frequency $\omega_{0}$. If the shift in $\omega_{0}$ is large, the resonance may be destroyed. We will focus on the QCD axion here. The axion field in the non-relativistic limit can be written as~\cite{correlation},
\begin{equation}
a(\vec{x},t)=\frac{1}{\sqrt{2m_{a}}}\left[e^{-im_{a}t}\psi(\vec{x},t)+c.c.\right] \, ,
\end{equation}
where $\psi(\vec{x},t)$ satisfies,
\begin{equation}
i\dot{\psi}=-\frac{1}{2m_{a}}\nabla^{2}\psi+\frac{\lambda}{8m_{a}^{2}}|\psi|^{2}\psi \, ,
\end{equation}
where $\lambda=-0.346m_{a}^{2}/f_{a}^{2}$ is the coupling of self-interaction for the QCD axion~\cite{precise}. A correction $\Delta\omega_{0}=\lambda n_{a}/(8m_{a}^{2})$ is introduced, where $n_{a}$ is the axion number density, leading to a shift in the resonant frequency. For axion clumps with non-uniform densities, the shift must be small enough so that the resonance could occur. Because resonance occurs near $\omega=m_{a}/2$, the resonant bandwidth of the axion field for fixed electromagnetic wave frequency is twice as large as the resonant bandwidth of electromagnetic wave $\Delta\omega=g_{a\gamma}\sqrt{{\rho_{a}}/{2}}$. Hence the requirement for resonance to occur is,
\begin{equation}
\frac{\lambda n_{a}}{8m_{a}^{3}}<\frac{2g_{a\gamma}}{m_{a}}\sqrt{\frac{\rho_{a}}{2}} \, ,
\end{equation}
\begin{equation}\label{eq:self}
\rho_{a}<2.62\times10^{9}\, {\rm g/cm^{3}}\left(\frac{g_{a\gamma}}{10^{-12}\, {\rm GeV^{-1}}}\right)^{2}\left(\frac{m_{a}}{10^{-2}\, {\rm eV}}\right)^{-2}\, ,
\end{equation}
where we have used eq.~(\ref{eq:massfa}). Axion clumps with densities as high as the limit in eq.~\eqref{eq:self} are irrelevant in our discussions. Dense axion stars may have such a high density, but it is potentially unstable due to the scattering of axions to a relativistic velocity~\cite{star}.

Eq.~(\ref{eq:self}) should be viewed as a rough estimate. In such a high density case, the non-relativistic approximation can be wrong. Gravitational effect and higher order terms in axion self-interaction may also be important. All these factors could possibly change the physical picture in a high density region and the result of eq.~(\ref{eq:self}) can completely change. For the low density case that we discuss in this work, eq.~(\ref{eq:self}) shows that self-interaction does not stop the resonance.
\subsection{Velocity}

If an axion clump reaches an equilibrium configuration with a zero angular
momentum, we can neglect the effect of velocity. However, if there is
a macroscopic velocity in the clump, which is possible for axion clumps that
are not in equilibrium, the Doppler effect changes the frequency
of electromagnetic waves, possibly leading to a breakdown
of the resonance. Axion clumps may also have a nonzero angular momentum, whose
effect was studied in ref.~\cite{resonance2}. The global motion of axion clumps in galaxies, however, has no effect. Only
the relative motion within an axion clump is important. From the point of view of field theory, these motions will introduce a rapid
spatial fluctuation term $\sim e^{ikx}$, which may affect the
resonance. This leads to a shift of the resonant frequency, proportional to the velocity $v$. The relative bandwidth in eq.~(\ref{eq:bandwidth}) is about $10^{-10}$ to $10^{-19}$ for an axion clump density from $0.1\, {\rm g/cm^{3}}$ to $10^{-19}\, {\rm g/cm^{3}}$, which means that the resonance is sensitive to
the macroscopic velocity. We must require the macroscopic velocity to be smaller than $10^{-10}c$ to $10^{-19}c$, which seems hard to achieve for axion
clumps not in equilibrium.

For the random motion of axions, the velocity changes the energy of axions,
$\omega_{0}$, but no spatial fluctuation terms like $e^{ikx}$ appear. The
system is static macroscopically. So the shift in the frequency is of order
$v^2$, much smaller than the effect of the macroscopic velocity.

\subsection{Parameter space for instability in axion clumps}

\begin{figure}[htbp]
  \centering
  \includegraphics[width=14cm]{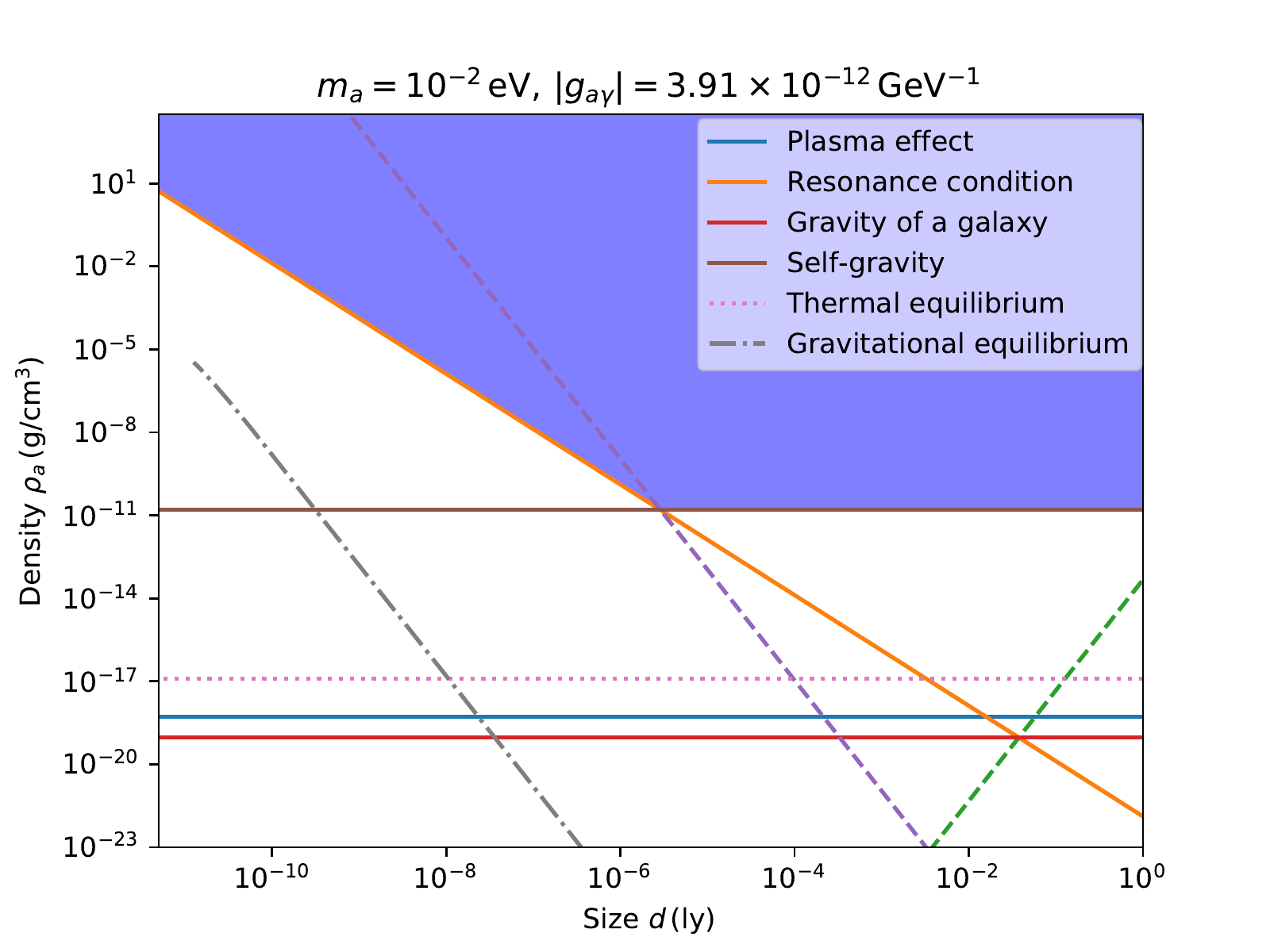}\\
  \caption{Instability region in the axion clump density-size parameter space,
assuming a uniform density with axion mass $m_{a}=10^{-2}\, {\rm eV}$ and an
axion-photon coupling $|g_{a\gamma}|=3.91\times10^{-12}\, {\rm GeV^{-1}}$.
Axion clumps in the blue shaded region has resonant instability. Plasma effect,
gravitational redshift from the galaxy and self-gravity, resonant condition
are used to constrain the parameter space where the resonant instability
occurs. We take the electron density $n_{e}=0.03\, {\rm cm^{-3}}$ and the radius of the equi-potential surface $R=10^{4}\, {\rm ly}$ here. Green and blue dashed lines are respectively the dividing lines
where the gravitational redshift of the galaxy and self-gravity are within
bandwidth $\Delta\omega=g_{a\gamma}\sqrt{\rho_{a}/2}$ for the whole axion
clump. The ``$\sech$'' approximation~\cite{clump} $\alpha(r)=\alpha_{0}\sech(r/R)$ is used, where
$\alpha(r)$ is amplitude of the axion field and $\alpha_{0}$ is a constant.
The density of equilibrium configurations is taken as the average density
within radius $R$. Thermal equilibrium could be reached by self-interaction
of axions if the axion clump density is higher than that indicated by the
pink dotted line. Note that axion clumps in equilibrium have no resonant instability here.}
  \label{img1}
\end{figure}

Resonant instability occurs only if an axion clump has a large enough density and size. Here, we expect there is an instability region in the density-size parameter space of axion clumps, which depends on axion parameters, including
the axion mass $m_{a}$ and the axion-photon coupling $g_{a\gamma}$. We will work with fixed axion parameters first. Typically, the QCD axion mass is given by the constraint on $f_{a}$ and the relic density of axionic dark matter, ranging from $10^{-5}\, {\rm eV}$ to $10^{-2}\, {\rm eV}$. We first take the axion mass $m_{a}=10^{-2}\, {\rm eV}$, which is indicated by stellar cooling
observations~\cite{cooling}, and $|g_{a\gamma}|=3.91\times10^{-12}\, {\rm
GeV^{-1}}$ in the KSVZ model for such a mass. The four constraints given by the resonant condition, the galaxy's gravity, the self-gravity and the plasma effect, are plotted in figure~\ref{img1}. The effect from the self-interaction is small and does not show up in the figure. The effect from velocity is not considered here. All the lines in figure~\ref{img1}, except the line of gravitational equilibrium states, assume a uniform axion clump density. The blue shaded region is unstable. Axion clumps in the blue shaded region will decay via stimulated emission to photons which lowers the density of axions.

\begin{figure}[htbp]
  \centering
  \includegraphics[width=14cm]{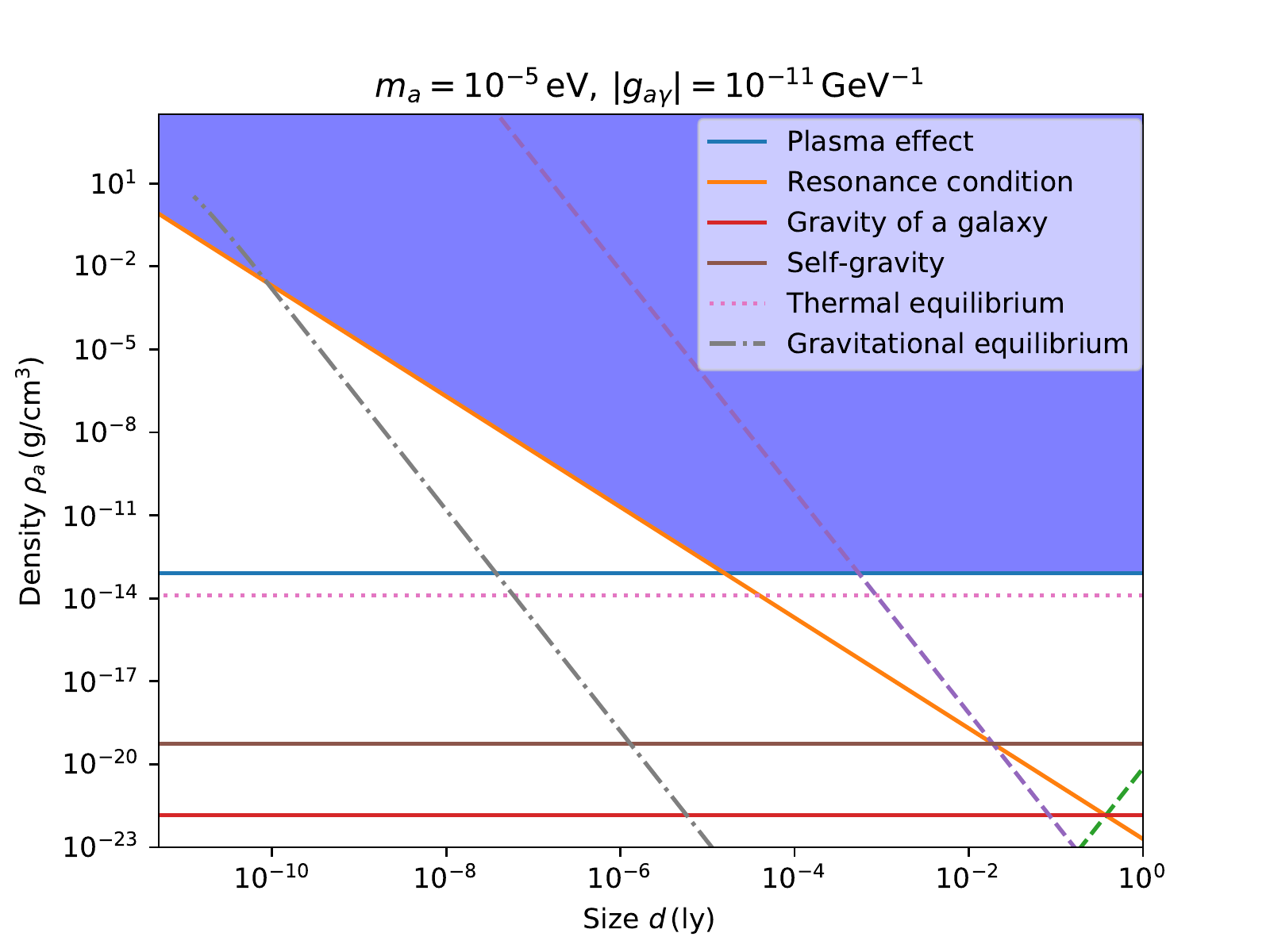}\\
  \caption{Similar as figure~\ref{img1}. Instability region in the axion clump density-size parameter space, assuming a uniform density with ALP mass $m_{a}=10^{-5}\, {\rm eV}$ and $|g_{a\gamma}|=10^{-11}\, {\rm GeV^{-1}}$. We still take the electron density $n_{e}=0.03\, {\rm cm^{-3}}$ and the radius of the equi-potential surface $R=10^{4}\, {\rm ly}$ here. Axion clumps in the blue  shaded region has resonant instability. Unlike in figure~\ref{img1}, here the dash-dotted line representing equilibrium configurations crosses the blue region where the resonance can occur.}
  \label{img2}
\end{figure}

The two dashed lines are constraints in eqs.~(\ref{eq:grav1}) and (\ref{eq:grav2}) from the gravitational redshift. If
these two constraints are satisfied, the resonant instability will occur in the whole axion clump. However, if some regions of an axion clump experience resonant instability, the clump is still unstable while eqs.
(\ref{eq:grav1}) and (\ref{eq:grav2}) are not satisfied. As illustrated in section~\ref{subsec:gravity}, the resonant instability still occurs if we increase the size of axion clump while keeping the density unchanged. Hence if an axion clump reaches the critical density, which equals to the intersection point of the line of the resonance condition and the dashed
line, resonance is allowed. So the constraints from gravitational redshift are represented by two horizontal solid lines shown in figure~\ref{img1}.

A detailed numerical calculation shows some modifications to the constraints from plasma effect and gravitational redshift. For the resonant instability to occur, resonance condition (\ref{eq:resonantcon}) must be satisfied and the shift of electromagnetic wave frequency should be within the bandwidth (\ref{eq:bandwidth}). However, the resonance condition is obtained for the central frequency $\omega=m_{a}/2$, not for the slightly shifted frequencies. For shifted frequencies, resonance requires higher axion clump densities. So if the left-hand side of eq.~(\ref{eq:resonantcon}) for an axion clump is only a little bit larger than $\pi/2$ for the uniform density case, resonance is easily destroyed by the frequency shift, even if the frequency shift is within the bandwidth (\ref{eq:bandwidth}). For resonant instability to occur in such cases, higher densities are required compared with previous estimations. So in figure~\ref{img1} and \ref{img2}, the lines of plasma effects and gravitational redshift should bend upwards near the orange line for the resonance condition.

The line that represents the gravitationally balanced configurations is also plotted~\cite{clump}. A hyperbolic approximation, $\alpha(r)=\alpha_{0}\sech(r/R)$, is used, where $\alpha(r)$ is the amplitude of the axion field and $\alpha_{0}$ is a constant. The density of equilibrium configurations shown in figure~\ref{img1} and \ref{img2} is taken as the average density within radius $R$, while the size is taken as $d=2R$. Sometimes those axion clumps in equilibrium are called dilute axion stars. Note that the line of gravitational equilibrium states ends at the high density side, which means a maximally allowed density exists. For higher densities, dilute axion stars may collapse to form dense axion stars, where self-interaction is important. These dense axion stars are probably unstable due to radiation of relativistic axions~\cite{star}. So we focus on the dilute axion star case. It is shown from figure~\ref{img1} that the resonant instability will not occur for dilute axion stars for axion parameters $m_{a}=10^{-2}\, {\rm eV}$ and $|g_{a\gamma}|=3.91\times10^{-12}\, {\rm GeV^{-1}}$.

We plot in figure~\ref{img2} for another ALP parameters
$m_{a}=10^{-5} \, {\rm eV}$, indicated by the dark matter relic density from misalignment mechanism, and $g_{a\gamma}=10^{-11} \, {\rm GeV}^{-1}$,
whose coupling to photons is much larger than the QCD axion model
for such a mass. Note that the axion-photon coupling we take here is still unconstrained by the CAST~\cite{CAST2}. As shown in figure~\ref{img2}, plasma effect is more important than gravitational redshift for such axion parameters. Resonant instability is possible in this case for equilibrium configurations if the density of the dilute axion star is high enough.

If such a resonant instability occurs in an axion clump, a strong radiation may be observable~\cite{clump2,axlaser}. The electromagnetic wave
emitted will be monochromatic. Although astrophysical processes may give a
broader spectrum of such sources, the peak frequency $\omega=m_{a}/2$,
usually in the microwave or infrared band, will still be much
more prominent than other frequencies, which is possible to identify. The total
energy released should be the same order of the total mass of axion clump. The time duration of resonant instability is roughly several times of the light-crossing timescale. If axion clumps come close to a neutron star with a strong magnetic field, similar resonant conversion process also occurs, and is possibly related to fast radio burst~\cite{axFRB}. If such explosions are observed, it may be a strong evidence for the axionic dark matter.

We now consider an axion clump in equilibrium, in which case the density has an upper limit. For an axion clump with a higher density, or equivalently a higher mass, the system becomes unstable and will collapse. If we use the ``sech'' approximation $\alpha(r)=\alpha_{0}\sech(r/R)$, there is a relation between the central density $\rho_{a}$ and the characteristic radius $R$ for equilibrium configurations~\cite{clump}. We define a dimensionless radius $\tilde{R}=m_{a}f_{a}\sqrt{G}R$ and a rescaled particle number $\tilde{N}=m_{a}^2N\sqrt{G}/f_{a}$. A relation for equilibrium configurations is~\cite{clump},
\begin{equation}
\tilde{R}=\frac{a\pm\sqrt{a^2-3bc\tilde{N}^2}}{b\tilde{N}} \, ,
\end{equation}
\begin{align}
  a=& \frac{12+\pi^2}{6\pi^2} \, , \\
  b=&\frac{6\left[12\zeta(3)-\pi^2\right]}{\pi^4} \, , \\
  c=&\frac{\pi^2-6}{8\pi^5} \, ,
\end{align}
where we will take the ``$+$'' sign because the ``$-$'' sign corresponds to unstable solutions. A dilute axion clump has a minimum radius and hence a maximum density when the number in the squared root is zero, yielding a minimum radius
\begin{equation}
\tilde{R}=\frac{a}{b\tilde{N}} \, .
\end{equation}
The corresponding central density is~\cite{clump}
\begin{equation}
\rho_{a}=\frac{3Nm_{a}}{\pi^3 R^3} \, .
\end{equation}
Resonant instability is most likely to occur for dilute axion clumps with
the highest possible density. So we only need to check whether configurations with the highest density are resonantly unstable. It turns out that only the resonance condition is relevant to determine the stability. The resonance condition for the ``$\sech$'' ansatz
is,
\begin{equation}
g_{a\gamma}R\sqrt{\frac{\rho_{a}}{2}}>1 \, .
\end{equation}
The relation between $g_{a\gamma}$ and $f_{a}$ is (for $C=1$),
\begin{equation}
g_{a\gamma}=\frac{\alpha_{\rm EM}}{2\pi f_{a}}C_{a\gamma} \, .
\end{equation}
Hence, the resonance condition becomes,
\begin{equation}
C_{a\gamma}>443.6 \, .
\end{equation}

\begin{figure}[htbp]
  \centering
  \includegraphics[width=14cm]{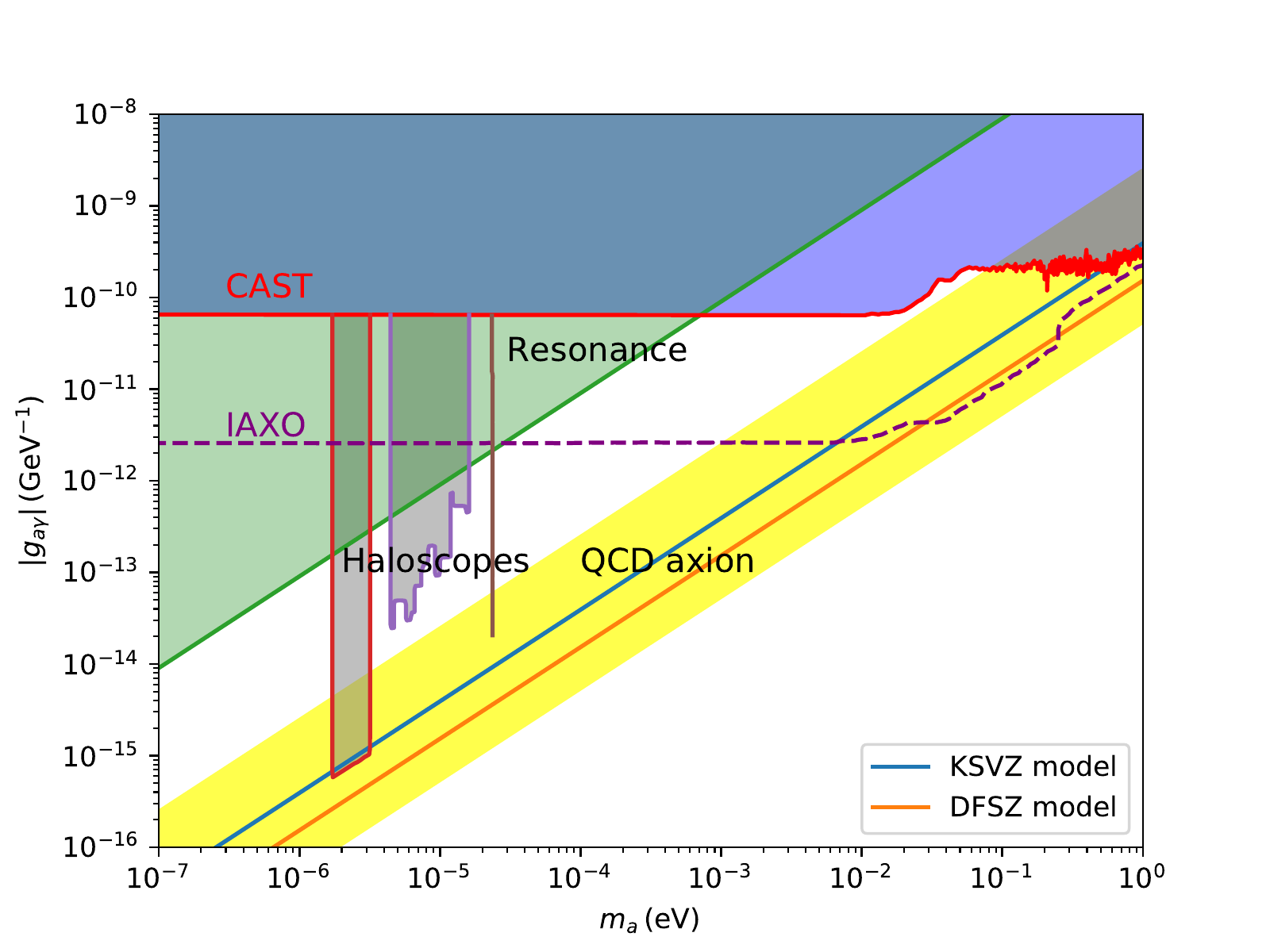}\\
  \caption{Instability region in the parameter space of the axion-photon coupling versus the axion mass. If axion parameters are in the green shaded
region, it is possible for the resonant instability to occur for dilute axion clumps in equilibrium. Constraints from CAST and haloscopes observations, and the planned IAXO detector are plotted~\cite{CAST2}. QCD axion models are in
the yellow shaded region determined by $0.25<C_{a\gamma}<12.75$~\cite{axionwin}.}
  \label{img3}
\end{figure}

The result applies to an axion clump in equilibrium with a density profile $\alpha(r)=\alpha_{0}\sech(r/R)$, which is a good approximation to numerical results~\cite{clump}. To plot the result in the axion parameter space, we must suppose that eq.~(\ref{eq:massfa}) is satisfied. The axion mass $m_{a}$ versus the coupling $g_{a\gamma}$ parameter space is shown in figure~\ref{img3}. If axion parameters are in the green shaded region, resonance will occur for dilute axion stars with the highest density allowed. While for those parameters outside the green shaded region, for example parameters in most QCD axion models, no resonant instability occurs for axion clumps in equilibrium. This result is consistent with previous results in refs.~\cite{clump2,resonance2,axFRB,axlaser}. The green shaded region should not be considered as a constraint on axion parameters if such resonance is absent, because we do not know whether dilute axion stars with the highest possible density exist in the Universe. Other constraints from CAST and observations of haloscopes, and the planned IAXO detector are also shown~\cite{CAST2}.

\section{Discussion}

Axions are an excellent dark matter candidate. If dark matter is mainly composed of axions, we would expect that they form miniclusters or clumps under the effect of gravity. Some axions may have a relatively high velocity dispersion~\cite{mini} and distribute outside these axion clumps, making it possible to detect axionic dark matter directly. Axion clumps, on the other hand, are hard to be observed, considering their low density and small mass. However, if resonant instability occurs in these axion clumps, they may
have a chance to be observed. The results in section
\ref{sec:instability} show that for QCD axion clumps that reach
equilibrium configurations, no resonant instability occurs except for a relatively large $g_{a\gamma}$. If the axion clump does not reach equilibrium configurations, resonant instability is possible for axion clumps in the blue region in figures~\ref{img1} and \ref{img2}. However, the conclusion is still a bit premature due to the uncertainty of the macroscopic velocity in these clumps.

Axion clumps may form BEC. The occupancy number of
dark matter axions is as large as $10^{26}$~\cite{correlation} or even
higher in axion clumps. But to form BEC, dark matter axions must reach
thermal equilibrium. For a usual axion self-interaction~\cite{precise},
$\mathcal{L}_{\rm int}=-\lambda a^{4}/4!$ with
$\lambda=-0.346m_{a}^{2}/f_{a}^{2}$, the relaxation rate is
$\Gamma_{\lambda}\sim|\lambda|n_{a}/(4m_{a}^{2})$. If an axion clump reaches
thermal equilibrium by self-interactions, we require at least
$\Gamma_{\lambda}>H_{0}$, where $H_{0}$ is the Hubble parameter at present.
The requirement of thermal equilibrium gives a lower bound on the axion
density,
\begin{equation}\label{eq:thermal}
\rho_{a}>1.26\times10^{-17} \, {\rm g/cm^{3}} \left(\frac{m_{a}}{10^{-2} \, {\rm eV}} \right)^{-1} \, .
\end{equation}
The result is shown as pink dotted lines in figures~\ref{img1} and \ref{img2}. If the density of axionic dark matter is higher than the density in eq.~(\ref{eq:thermal}), thermal equilibrium could be reached by self-interaction. However the result above only takes axion self-interactions into account. There is some controversy on whether the gravitational interaction can thermalize
axion clumps~\cite{relaxation}, which we will not discuss in detail here.
If axion clumps form BEC, we would expect that almost all axions occupy
the ground state, in which case only the ground state needs to be dealt
with. Otherwise, the effective number of axions in the resonant process
might be reduced.

Resonant instability in axion clumps in itself does not give a strong
constraint on the axion-photon coupling $g_{a\gamma}$. In a scale larger than one light year, the gravitational redshift effect could stop the resonance completely. It is unlikely that axionic dark matter can have a density
within the resonant instability region with a size much larger than
one light year. This is because that the density required for
resonant instability is at least 6 (or 14, for figure \ref{img1}) orders of magnitude higher than the Solar local dark matter density, $0.3 \, {\rm GeV/cm^{3}}$. Although some enhancement may be present, due to the small bandwidth, which is unresolvable in observations, no obvious enhancement will occur in reality.

The discussion is similar for scalar ALPs, whose coupling to photons is
$-\frac{1}{4}g_{a\gamma}aF_{\mu\nu}F^{\mu\nu}$. Eq.~(\ref{eq:fieldeq1}) is slightly modified in this case. For
non-relativistic ALPs, $\nabla a$ is negligible and we can drop higher order
terms. The only change is that the $\nabla\times\vec{A}$ term
in eq.~(\ref{eq:fieldeq1}) becomes $\partial\vec{A}/\partial t$.
Solutions still show a resonant instability at $\omega=m_{a}/2$ with the
same exponential growth rate (Floquent exponent). So the overall picture
should be similar.

\acknowledgments

Z. Wang and L.-X. Li were supported by the National Natural Science Foundation of China (Grant No. 11973014). L. Shao was supported by the National Natural Science Foundation of China (Grant Nos. 11975027 and 11991053) and the Young Elite Scientists Sponsorship Program by the China Association for Science and Technology (Grant No. 2018QNRC001).

\normalem
\bibliographystyle{unsrt}
\bibliography{refs}

\end{document}